\documentclass[%reprint,
superscriptaddress,
reprint,
 amsmath,amssymb,
 aps,
]{revtex4-2}

\usepackage{color}

\usepackage{graphicx}% Include figure files
\usepackage{dcolumn}% Align table columns on decimal point
\usepackage{bm}% bold math
\begin{document}

\preprint{APS/123-QED}

\title{Mathematical modeling for the synchronization of two interacting active rotors}

\author{Hiroyuki Kitahata}
\email{kitahata@chiba-u.jp}
\affiliation{Department of Physics, Graduate School of Science, Chiba University, Chiba 263-8522, Japan
}

\author{Yuki Koyano}
\email{koyano@garnet.kobe-u.ac.jp}
\affiliation{Department of Human Environmental Science, Graduate School of Human Development and Environment, Kobe University, Kobe 657-0011, Japan}

\date{\today}% It is always \today, today,

\begin{abstract}
We investigate the synchronization of active rotors. A rotor is composed of a free-rotating arm with a particle that releases a surface-active chemical compound. It exhibits self-rotation due to the surface tension gradient originating from the concentration field of the surface-active compound released from the rotor. In a system with two active rotors, they should interact through the concentration field. Thus, the interaction between them does not depend only on the instantaneous positions but also on the dynamics of the concentration field. By numerical simulations, we show that in-phase and anti-phase synchronizations occur depending on the distance between the two rotors. The stability of the synchronization mode is analyzed based on phase reduction theorem through the calculation of the concentration field in the co-rotating frame with the active rotor. We also confirm that the numerical results meet the prediction by theoretical analyses.

\end{abstract}

\maketitle

\section{Introduction}

Self-propelled particles have been intensively investigated for decades both experimentally and theoretically~\cite{Michelin,Ohta_JPSJ}. Motions of living organisms such as cells, bacteria, fish, birds, and insects attract much interest as examples of the self-propulsion. For such motions, several mechanisms on the motion are suggested, e.g., hydrodynamic interaction due to the surface deformation~\cite{Klindt_2017,Uchida_JPSJ}, momentum exchange with the substrate~\cite{Sens_PNAS,Tarama_JPSJ}, and the tactic motions~\cite{Budrene1995DynamicsOF,Shoji_JPSJ,Nakajima2014RectifiedDS}. As for the tactic motions, they are classified into several types such as chemotaxis, phototaxis, mechanotaxis, geotaxis, and so on. Here, we focus on the chemotactic motion, in which the direction of motion is determined by the concentration gradient. For positive and negative chemotaxes, the object moves in the positive and negative directions of the gradient of the concentration field, respectively. If the object releases a chemical compound around itself and exhibits negative chemotaxis, then the rest state where the object stands still can become unstable since the object is likely to move away from the original position with higher concentration. The motion can be sustained since the particle motion can keep the anisotropy in the concentration field around the particle. This is one of the mechanism for the self-propulsion with the taxis~\cite{mikhailov2002cells}.

An experimental example for such self-propulsion with negative chemotaxis is a camphor particle floating at a water surface. The camphor particle releases the camphor molecules at the water surface, and the molecules reduce the surface tension. The object is pulled toward the region with higher surface tension reflecting lower camphor concentration, which can be understood as negative chemotaxis~\cite{Tomlinson1862,Strutt_rspl1890,Nakata_Langmuir1997,PCCP_camphor_review2015,Nakata-book,Boniface}. Several types of active rotors, or self-propelled rotors, were recently reported using camphor or some other chemicals with surface activity~\cite{Ei2018,Shimokawa_JPSJ,Koyano_PRE.96.012609,Chaos_Koyano2019,Sharma_PRE.99.012204,Sharma_PRE.101.052202,Sharma_PRE.103.012214,Sharma_PRE.105.014216,Sharma_PRE.106.024201,MOROHASHI2019104,Frenkel_APL,Frenkel2019,Koyano_imperfection}. For example, an elliptic camphor paper can rotate around an axis penetrating a hole at the center of the paper~\cite{Ei2018,Shimokawa_JPSJ}. We also reported a camphor rotor, in which a plastic plate with two or more camphor particles can freely rotate around an axis penetrating at the center of the plate~\cite{Koyano_PRE.96.012609,Chaos_Koyano2019}. Some other types of rotors using camphor or other surface-active compounds have also been reported.

As theoretical approaches for the camphor particle motion, the reaction-diffusion equation for camphor concentration coupled with the Newtonian equation for a camphor particle motion has often been adopted~\cite{HayashimaJPCB2001,Nagayama_PhysicaD,PCCP_camphor_review2015,Nakata-book}. We previously discussed the bifurcation of a camphor rotor by considering the time evolution for the camphor concentration coupled with the Newtonian equation for the rotation of the arm attached with camphor particles~\cite{Koyano_PRE.96.012609,Chaos_Koyano2019}. We theoretically derived the simplified ordinary differential equation on the angle of the camphor rotor using the perturbation method and showed that the self-propelled rotation emerges through supercritical pitchfork bifurcation by changing the friction coefficient as a bifurcation parameter. Sharma and co-authors recently reported the results of experiments and numerical simulation based on the simple mathematical model. In their experimental system, the two rotors made of rectangular camphor papers were closely located, and they observed the desynchronization and anti-phase synchronization in our definition, i.e., the particles alternately come close to the other rotor. In their numerical simulation, they assume that a camphor rotor has an intrinsic angular velocity and interacts with the other rotor by a Yukawa-type potential depending on their relative position. They succeeded in reproducing their experimental results~\cite{Sharma_PRE.99.012204}. They also investigated the interaction between the multiple rotors and reported many interesting states such as synchronous, quasiperiodic, and chaotic states~\cite{Sharma_PRE.101.052202,Sharma_PRE.103.012214,Sharma_PRE.105.014216,Sharma_PRE.106.024201}.

As mentioned above, a single rotor composed of the particle with a surface-active compound rotates due to the surface tension gradient at the water surface, and the surface tension is a function of the concentration of the compound. In the typical experiments with rotors with camphor particles, the rotation period of the rotors is of the order of 1~s and the effective diffusion length of the concentration field is several tens of millimeters. Thus, the characteristic time for the chemical compound to diffuse to the other rotor is comparable to the rotation periods. Therefore, we consider that the interaction through the concentration field is important when the distance between rotors is large. Thus, here we discuss the dynamics of camphor rotors including the time evolution of the concentration field. We found that the in-phase or anti-phase synchronization mode occurs depending on the distance between the rotors. In addition, we succeeded in the mathematical analysis based on the phase description.

Our manuscript is constructed as follows: We first describe the mathematical model for the active rotors in Sec.~\ref{sec:modelling} and then show the results of numerical simulation based on the model in Sec.~\ref{sec:numerical}. Then, we perform the theoretical analyses using the phase reduction. The procedure and results of the analyses are described in Sec.~\ref{sec:analysis}. Then, we discuss the validity of the phase description by directly calculating the phase coupling function using numerical simulation in Sec.~\ref{sec:discussion}. Finally, we summarize the results and show a possible extension of our model in Sec.~\ref{sec:conclusion}.

\section{Mathematical Model \label{sec:modelling}}

We construct a mathematical model for a system with symmetric active rotors, which can rotate in either clockwise or counterclockwise direction, based on the previous studies~\cite{Nagayama_PhysicaD,PCCP_camphor_review2015,Nakata-book}. We mainly consider a two-rotor system, but also consider a single-rotor one to clarify the characteristics of a composing rotor. Our model comprises the time-evolution equation for the configurations of the rotors with particles (camphor particles) that release a surface-active compound (camphor molecules), and that for the concentration of a surface-active compound.

\begin{figure}
\includegraphics{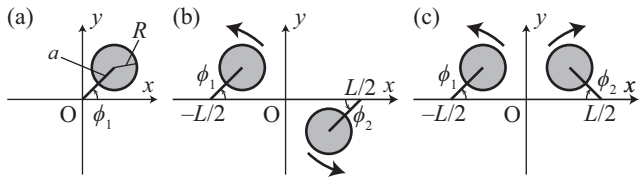}
\caption{\label{fig1} Schematic illustration for the active rotors and definition of phases. (a) A single rotor. (b) Two coupled rotors which rotate in the same direction (counterclockwise). (c) Two coupled rotors, which rotate in different directions. The first (left) and second (right) ones rotate counterclockwise and clockwise, respectively.}
\end{figure}

The $i$th particle can move along a circle with a radius of $a$ and the center position of $\bm{\ell}_i$. Therefore, the particle position can be described only by using one variable $\phi_i$, which is called the phase of the $i$th particle. We define the origin and positive direction of each phase as schematically shown in Fig.~\ref{fig1}. 
For a single-rotor system as in Fig.~\ref{fig1}(a), we set $\bm{\ell}_1 = \bm{0}$ and thus we can express the position of the particle position $\bm{r}_1$ as
\begin{align}
    \bm{r}_1 = \bm{\ell}_1 + a \bm{e}(\phi_1) = a \bm{e}(\phi_1). \label{eq_r1}
\end{align}
For a two-rotor system as in Figs.~\ref{fig1}(b) and (c), we set $\bm{\ell}_1 = - (L/2)\bm{e}_x$ and $\bm{\ell}_2 = (L/2)\bm{e}_x$.
For the symmetric expression between the first and second rotors on the time evolution of each phase, the origins of the phases are set so that $\phi_i = 0$ corresponds to the direction toward the center of the other rotor, and positive directions of the phase are set as the rotation direction of each rotor.
That is to say, the positions of the two particles $\bm{r}_1$ and $\bm{r}_2$ are expressed using the phases $\phi_1$ and $\phi_2$ as
\begin{align}
    \bm{r}_1 = \bm{\ell}_1 + a \bm{e}(\phi_1) = -\dfrac{L}{2} \bm{e}_x + a \bm{e}(\phi_1),
\end{align}
\begin{align}
    \bm{r}_2 = \bm{\ell}_2 - a \bm{e}( \pm \phi_2) = \dfrac{L}{2} \bm{e}_x - a \bm{e}( \pm \phi_2). \label{eq_rotor2}
\end{align}
In Eq.~\eqref{eq_rotor2}, the positive and negative signs correspond to the rotation in the same direction (Fig.~\ref{fig1}(b)) and in the opposite direction (Fig.~\ref{fig1}(c)), respectively. Here, we set the Cartesian coordinates so that the origin meets the midpoint of the centers of the two rotors and the line connecting the centers of the two rotors meets the $x$ axis. The unit vector in the $x$ and $y$ axes is set as $\bm{e}_x$ and $\bm{e}_y$, respectively, and $\bm{e}(\theta)$ is a unit vector in the direction of $\theta$, i.e., $\bm{e}(\theta) = \cos \theta \bm{e}_x + \sin \theta \bm{e}_y$. 

The time evolution equation for the rotor is obtained based on the Newtonian equation with the overdamped scheme. That is to say, the equation is written as
\begin{align}
    \eta_i A \frac{d\bm{r}_i}{dt} = \eta_i A \bm{v}_i = \bm{F}_{u,i} + \bm{F}_{c,i}, \label{eq_motion}
\end{align}
where $\bm{v}_i$ is the velocity of the particle composed of the $i$th rotor, $\eta_i$ is the friction coefficient per area for the $i$th rotor, and $A$ is the area of the particle. The velocity $\bm{v}_i$ is expressed using the phase $\phi_i$ as
\begin{align}
\bm{v}_1 = a \frac{d\phi_1}{dt} \bm{e}\left(\phi_1 + \frac{\pi}{2}\right),
\end{align}
\begin{align}
\bm{v}_2 = a \frac{d\phi_2}{dt} \bm{e}\left(\phi_2 \pm \frac{\pi}{2}\right),
\end{align}
where the positive and negative signs correspond to the cases with the same and opposite rotation directions, respectively.
$\bm{F}_{ u,i}$ and $\bm{F}_{c,i}$ are the force exerted on the $i$th particle due to the surface tension gradient and the constraint force in the direction of $\bm{e}(\phi_i)$. The force due to the surface tension gradient is expressed using area integration as
\begin{align}
    \bm{F}_{u,i} 
    =& \oint_{\partial \Omega_i} \left[ - \Gamma u \bm{n}\right] d\ell \nonumber \\
    =& \iint_{\Omega_i} \left[ - \Gamma \nabla u  \right] dA \nonumber \\
    =& - \Gamma \iint_{\mathbb{R}^2} \left( \nabla u \right) S\left(\bm{r} - \bm{r}_i\right) dA,
\end{align}
where $u$ is the concentration field of the surface-active chemical compound, $\Omega_i$ is the region of the particle composing the $i$th rotor, $\partial \Omega_i$ is the periphery of $\Omega_i$, and $\bm{n}$ is the outward normal unit vector at the particle periphery. $dA$ and $d\ell$ are the area and line elements. The surface tension should be a decreasing function of the concentration of the surface-active compound.
Here, we assume a linearity between surface tension and the concentration, where the proportionality constant is $-\Gamma$. $S\left(\cdot \right)$ is a level function. In the numerical simulation, we adopt a smoothed step function which has values close to unity inside the particle and values close to zero outside of it, i.e.,
\begin{align}
    S(\bm{\ell}) = \left\{ \begin{array}{ll}
    1, & \left| \bm{\ell} \right| \leq R - \varepsilon, \\
    1 - \left( \left| \bm{\ell} \right| - R + \varepsilon \right)^2 / \left( 2 \varepsilon^2 \right),& R - \varepsilon < \left| \bm{\ell} \right| \leq R, \\
    \left( \left| \bm{\ell} \right| - R - \varepsilon \right)^2 / \left( 2 \varepsilon^2 \right), & R < \left| \bm{\ell} \right| < R + \varepsilon, \\
    0, & \left| \bm{\ell} \right| \geq R + \varepsilon,
    \end{array}\right.
\end{align}
where $R$ is the radius of a particle of the surface-active compound and $\varepsilon$ is a smoothing factor.
Then, we obtain the evolution equation by calculating the vector product of Eq.~\eqref{eq_motion} with $a\bm{e}(\phi_i)$ as
\begin{align}
    \eta_1 A a^2 \frac{d\phi_1}{dt} =&  \left( \bm{r}_1 - \bm{\ell}_1 \right) \times \bm{F}_{1} \nonumber \\
    =& a \bm{e}(\phi_1) \times \left(\bm{F}_{u,1} + \bm{F}_{c,1}\right) \nonumber \\
    =& a \bm{e}\left(\phi_1 + \frac{\pi}{2} \right) \cdot \bm{F}_{u,1}, \label{eq_rot1}
\end{align}
and
\begin{align}
   \eta_2 A a^2 \frac{d\phi_2}{dt} =& \pm \left( \bm{r}_2 - \bm{\ell}_2 \right) \times \bm{F}_{2} \nonumber \\
    =& \mp a \bm{e}(\pm \phi_2) \times \left(\bm{F}_{u,2} + \bm{F}_{c,2}\right) \nonumber \\
    =& \mp a \bm{e}\left(\pm \phi_2 + \frac{\pi}{2} \right) \cdot \bm{F}_{u,2} \nonumber \\
    =& a \bm{e}\left( \pm \left(\phi_2 - \frac{\pi}{2} \right) \right) \cdot \bm{F}_{u,2}. \label{eq_rot2}
\end{align}
Here, the upper and lower signs correspond to the rotation in the same rotation direction and the opposite rotation direction, respectively. The operator ``$\times$'' denotes the vector product in two dimensions, i.e., $\bm{\alpha} \times \bm{\beta} = \alpha_x \beta_y - \alpha_y \beta_x$ for $\bm{\alpha} = \alpha_x \bm{e}_x + \alpha_y \bm{e}_y$ and $\bm{\beta} = \beta_x \bm{e}_x + \beta_y \bm{e}_y$. In the calculation, we used that the constraint force $\bm{F}_{c,1}$ is parallel to $\bm{e}(\phi_1)$ and thus $\bm{F}_{c,1} \times \bm{e}(\phi_1) = 0$. We also used that $\bm{F}_{c,2}$ is parallel to $\bm{e}(\pm \phi_2)$ and thus $\bm{F}_{c,2} \times \bm{e}(\pm \phi_2) = 0$.

The dynamics of the concentration field is described as 
\begin{align}
    \frac{\partial u}{\partial t} = \nabla^2 u - u + \frac{1}{A}\sum_{i=1}^N  S(\bm{r} - \bm{r}_i), \label{eqRD}
\end{align}
where the first, second, and third terms on the right side correspond to the diffusion, evaporation, and supply of the surface-active chemical compound.  
$S(\bm{r} - \bm{r}_i) / A$ denotes the supply of the surface-active compound from the $i$th particle located at $\bm{r}_i$, and $N = 1$ for a single-rotor system and $N=2$ for a two-coupled rotor system.

It should be noted that we used the equations with dimensionless variables. The length, time, and concentration are normalized with the diffusion length, $\sqrt{D / \kappa}$, the characteristic time of sublimation, $1/\kappa$, and the ratio between the supply rate and sublimation rate, $f/\kappa$. Here, $D$ is the effective diffusion constant of the surface-active chemical compound~\cite{suematsu2014quantitative,JCP_HK_NY}, $\kappa$ is the sublimation rate, and $f$ is the supply rate of the compound for each disk.

\section{Numerical simulation \label{sec:numerical}}

Numerical simulation was performed based on the model in Sec.~\ref{sec:modelling}. For the numerical simulation, we adopted the Euler method for the dynamics of the rotors in Eqs.~\eqref{eq_rot1} and \eqref{eq_rot2} and with an explicit method for the dynamics of the concentration in Eq.~\eqref{eqRD}. The program code was prepared by ourselves in the C language. The calculation was performed in the region with $-X/2 \leq x \leq X/2$ and $-Y/2 \leq y \leq Y/2$. The Robin boundary condition $\nabla u\cdot\bm{n}_b + u = 0$ was adopted~\cite{Gustafson1998}, where $\bm{n}_b$ is the outward normal unit vector at the boundary. The initial condition was set as $u = 0$ in all of the calculation region. In order to stabilize the rotation direction, we fixed $d \phi_i / dt = 20$ for $0 \leq t \leq 1$. Due to the asymmetry in the concentration field around the particle maintained by the rotational motion, we succeeded in obtaining the stably rotating rotors. $\phi_1$ and $\phi_2$ at $t = 0$ were set to be $0$ and $\pi/2$, respectively, so that the convergence to the in-phase and anti-phase synchronization modes could be easily obtained.
The parameters were fixed as $\Gamma = 1$, $R = 0.1$, $a = 0.2$, and $\varepsilon = 0.025$. The spatial mesh and time step were $\Delta x = 0.025$ and $\Delta t = 0.0001$. The calculation region was  $X = 16$ and $Y = 10$. 

\begin{figure}
\includegraphics{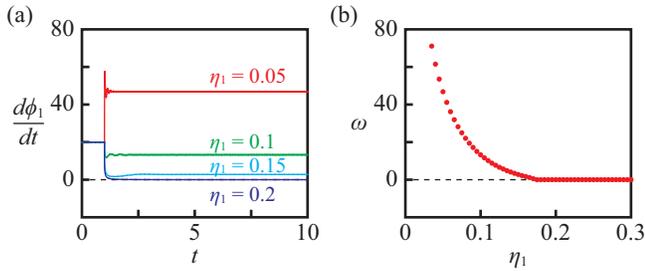}
\caption{\label{fig2} Numerical results for a single rotor. (a) Time series of $d\phi_1/dt$ for $\eta_1 = 0.05$ (red), $0.1$ (green), $0.15$ (cyan) and $0.2$ (blue). (b) The stable angular velocity $\omega$ depending on the friction coefficient $\eta_1$. A single rotor exhibited a rotation with a finite angular velocity for $\eta_1 < \eta_c \simeq 0.17$, while it stopped for $\eta_1 > \eta_c$.}
\end{figure}

First, we performed a numerical simulation for a single rotor. In Fig.~\ref{fig2}(a), we show the time series of $d\phi_1/dt$ for $\eta_1 = 0.05$, $0.1$, $0.15$, and $0.2$. For $\eta_1 = 0.05$, $0.1$, and $0.15$, $d\phi_1/dt$ converged to a finite positive value, while $d\phi_1/dt$ decayed to zero for $\eta_1 = 0.2$. We also confirmed that at $t = 10$, $d\phi_1/dt$ reached a steady value. Therefore, we defined a stable angular velocity $\omega$ as $d\phi_1/dt$ at $t = 100$. In Fig.~\ref{fig2}(b), the plot of $\omega$ against the friction coefficient $\eta_1$ is shown. The results suggest that a single rotor exhibited a stable rotation at a finite constant angular velocity for $\eta_1 < \eta_c$, while it stopped for $\eta_1 > \eta_c$, where $\eta_c \simeq 0.17$. This can be understood as a supercritical pitchfork bifurcation just as shown in the previous study~\cite{Koyano_PRE.96.012609}.

\begin{figure}
\includegraphics{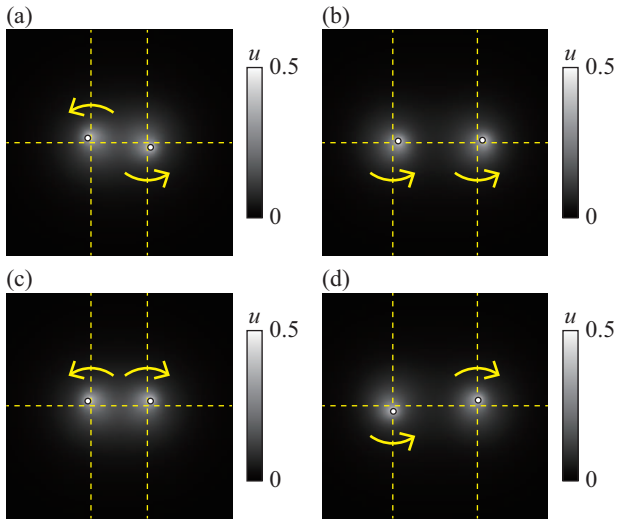}
\caption{\label{fig3}Snapshots representing the particle position and camphor concentration at $t = 1000$, at which the coupled rotors for $L = 2$ and $3$ reach the stable synchronization mode. (a) $L=2$ and (b) $L=3$ in the case with the same rotation direction. (c) $L = 2$ and (d)$L =3$ in the opposite rotation direction. The yellow arrows show the rotation direction. The cross points of the yellow dotted lines correspond to the centers of the rotors, $\bm{\ell}_1$ and $\bm{\ell}_2$.}
\end{figure}

\begin{figure}
\includegraphics{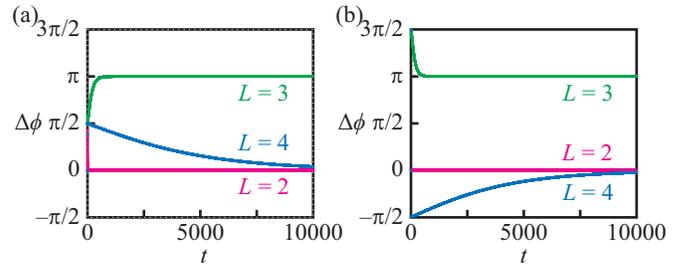}
\caption{\label{fig4} Numerical results for coupled rotors. Time series of $\Delta \phi$ are plotted for (a) the case with the same rotation direction and (b) the case with the opposite rotation direction. The distance between the two rotor centers $L$ was (a) $L = 2$ (red), (b) $L=3$ (green), and (c) $L = 4$ (cyan). In both cases, in-phase synchronization was observed for $L = 2$ and $4$, while anti-phase synchronization was observed for $L = 3$.}
\end{figure}

Then, we fixed $\eta_1 = \eta_2 = 0.1$ and calculated the behavior of the coupled system. We demonstrated the two cases: (i) the two rotors rotate in the same direction (cf. Fig.~\ref{fig1}(b)) and (ii) they do so in the opposite directions (cf. Fig.~\ref{fig1}(c)). In order to clarify the mode of synchronization, we detected the time $\tau^{(i)}_\mu$ at which the rotor $i$ passes through the line segment connecting $\bm{\ell}_1$ and $\bm{\ell}_2$ for the $\mu$th time. Then, the phase difference is defined as 
\begin{align}
\Delta \phi = 2\pi \frac{\tau^{(2)}_\mu - \tau^{(1)}_\nu}{\tau^{(1)}_{\nu+1} - \tau^{(1)}_\nu} ,
\end{align}
where $\mu$ and $\nu$ holds $\tau^{(1)}_\nu \leq \tau^{(2)}_\mu < \tau^{(1)}_{\nu+1}$. 

We changed $L$ and calculated the dynamics for the coupled rotors until $t = 10000$. The snapshots after the synchronized state becomes stable are shown for $L = 2$ and $3$ in each rotation direction in Fig.~\ref{fig3}.
The time evolution of $\Delta \phi$ is shown in Fig.~\ref{fig4}. In both cases with the same and opposite rotation directions, the in-phase synchronization ($\Delta \phi = 0$) was observed for $L = 2$ and $4$, while the anti-phase synchronization ($\Delta \phi = \pi$) was observed for $L = 3$. The relaxation time to the synchronized state was intensively dependent on the distance $L$ between the two rotors.

\begin{figure}
\includegraphics{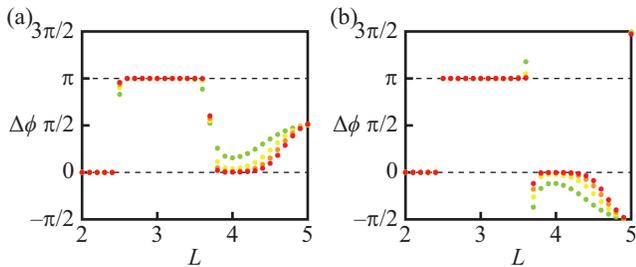}
\caption{\label{fig5} Numerical results for coupled rotors. The phase differences $\Delta \phi$ at $t = 5000$ (light green), $10000$ (yellow), $15000$ (orange), and $20000$ (red) are plotted in (a) the case with the same rotation direction and (b) the case with the opposite rotation direction. In-phase synchronization ($\Delta \phi = 0$) and anti-phase 
synchronization ($\Delta \phi = \pi$) alternate with an increase in $L$.}
\end{figure}

In order to clearly show the dependence of the synchronization mode on the distance $L$, we plotted $\Delta \phi$ at $t = 5000$, $10000$, $15000$, and $20000$ against $L$ in Fig.~\ref{fig5}. It clearly exhibits that the in-phase and anti-phase synchronization alternates with an increase in $L$. The preferable synchronization mode was almost the same in both cases. In Fig.~\ref{fig5}, the phase difference was not converged for the region greater than 4 and those close to the transition points between the in-phase and anti-phase synchronization. This should be because in these regions the interaction between the two rotors was so small that it takes much time to reach the stable synchronized state of the rotors. In such a case, synchronization cannot be observed in actual systems due to the inevitable fluctuation.

\begin{figure}
\includegraphics{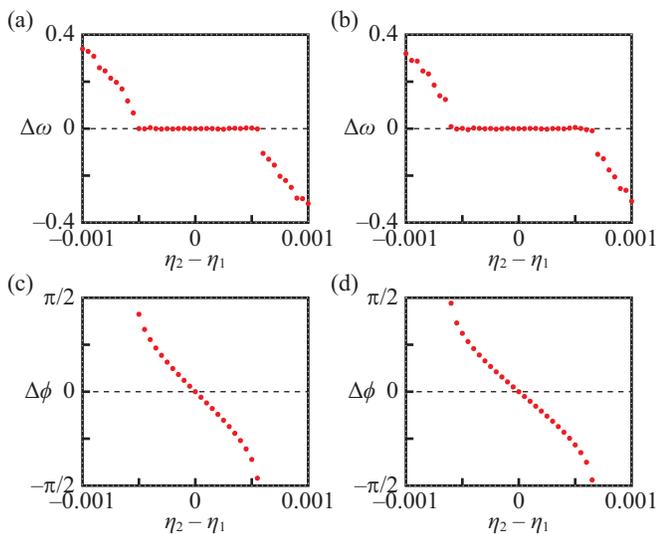}
\caption{\label{fig_dif} Numerical results on the synchronization for two coupled rotors with different intrinsic angular velocities by setting $\eta_1 \neq \eta_2$ with $L = 2$. Here, we fixed $\eta_1 = 0.1$ and varied $\eta_2$. (a), (b) The difference $\Delta \omega$ in the averaged angular velocities depending on $\eta_2 - \eta_1$ in the cases with (a) the same and (b) the opposite rotation directions. (c), (d) Phase difference for $\Delta \phi$ when the two rotors synchronized ($\left|\Delta \omega \right| < 0.01$) depending on $\eta_2 - \eta_1$ in the cases with (c) the same and (d) the opposite rotation directions.}
\end{figure}

In the above paragraphs, we only showed the numerical simulation results on the synchronization of the two identical rotors. In the synchronization of nonlinear oscillators, it is known that the synchronization can be observed for the two oscillators with slightly different intrinsic angular velocities~\cite{Pikovski,RMP_synchronization}. Thus, we considered the two slightly different rotors by fixing $\eta_1 = 0.1$ and varying $\eta_2$ when $L = 2$. We measured the averaged angular velocities $\omega_1$ and $\omega_2$ over $100 < t < 200$ and calculated the difference between them $\Delta \omega = \omega_2 - \omega_1$. The plots of $\Delta \omega$ against $\eta_2 - \eta_1$ are shown in Figs.~\ref{fig_dif} (a) and (b) in the case with the same and opposite rotation directions, respectively. We observed the synchronization ranges in $-0.0005 \lesssim \eta_2 - \eta_1 \lesssim 0.0005$, where $\Delta \omega = 0$. We also measured the phase difference $\Delta \phi = \phi_2 - \phi_1$ when the two rotors synchronized. As shown in Figs.~\ref{fig_dif}(c) and (d), the phase differences $\Delta \phi$ decreased with an increase in $\eta_2 - \eta_1$, and they are 0 when $\eta_2 = \eta_1$. This means that perfect in-phase synchronization was realized for the two identical rotors, while in-phase synchronization with a slight phase shift was realized for the two rotors with slightly different intrinsic angular velocities. These results suggest that the behavior of the coupled two-rotor system is translated in terms of the synchronization of the two nonlinear oscillators.

\section{Theoretical analysis \label{sec:analysis}}

In order to discuss the mechanism of the alternation of the stable synchronization modes depending on $L$, we perform the theoretical analysis to discuss the synchronization of the two active rotors based on the phase reduction method. Hereafter, we only consider the case with two identical rotors, i.e., $\eta_1 = \eta_2 = \eta$. The model in Eqs.~\eqref{eq_r1}--\eqref{eqRD} is used, but a point source is adopted, i.e. $R \to +0$ as 
 \begin{align}
    \frac{\partial u}{\partial t} = \nabla^2 u - u + \sum_{i=1}^N \delta(\bm{r} - \bm{r}_i), \label{eqRD_delta}
\end{align}
in the place of Eq.~\eqref{eqRD}, where $\delta(\cdot )$ is the Dirac's delta function.

Since the time-evolution equation for the concentration field $u$ in Eq.~\eqref{eqRD} is linear, $u$ is described as the summation of $u_1$ and $u_2$, which originate from the supply from the particles 1 and 2, respectively. That is to say,
\begin{align}
    u = u_1 + u_2,
\end{align}
where
\begin{align}
    \frac{\partial u_i}{\partial t} = \nabla^2 u_i - u_i + \delta(\bm{r} - \bm{r}_i),
\end{align}
for $i = 1,2$.

As for the time evolution of $\phi_i$, Eqs.~\eqref{eq_rot1} and \eqref{eq_rot2} give
\begin{align}
    \eta a^2 \frac{d\phi_1}{dt} 
    =& a \bm{e}\left(\phi_1 + \frac{\pi}{2} \right) \cdot \frac{1}{A} \bm{F}_{u,1}, 
\end{align}
and
\begin{align}
    \eta a^2 \frac{d\phi_2}{dt} 
    =& a \bm{e}\left( \pm \left(\phi_2 - \frac{\pi}{2} \right) \right) \cdot \frac{1}{A} \bm{F}_{u,2}.
\end{align}
The force originating from the surface tension $\bm{F}_{u,i}$ is also decomposed into two terms:
\begin{align}
    \bm{F}_{u,i} = \bm{F}_{u,i,1} + \bm{F}_{u,i,2},
\end{align}
in the same way as in the concentration $u$.
Under the point-source approximation, 
\begin{align}
\frac{1}{A} \bm{F}_{u,i,j} 
    =& - \frac{\Gamma}{A} \iint_{\mathbb{R}^2} \left( \nabla u_j \right) \delta\left(\bm{r} - \bm{r}_i\right) dA, \nonumber \\
    \to & -\Gamma \left.\nabla u_j\right|_{\bm{r} = \bm{r}_i}, \label{eq_Fij}
\end{align}
for $i \neq j$. It should be noted that the expression in Eq.~\eqref{eq_Fij} for $i = j$ does not hold since the force $\bm{F}_{u,i,i}/A$ shows the logarithmic divergence and we should introduce a small positive value corresponding to the particle radius~\cite{Koyano_PRE.96.012609,Koyano_2D}.
Nevertheless, from the physical point of view, the $i$th rotor should rotate with a constant angular velocity for $t \to \infty$ when it is driven only by $\bm{F}_{u,i,i}$. We define the terminal angular velocity to be $\omega$ and then the following equations hold:
\begin{align}
    \frac{1}{\eta A a} \bm{e}\left( \phi_1 + \frac{\pi}{2}\right) \cdot \bm{F}_{u,1,1} = \omega, \label{rot_av1}
\end{align}
for rotor 1, and
\begin{align}
    \frac{1}{\eta A a} \bm{e}\left( \pm\left( \phi_2 - \frac{\pi}{2} \right)\right) \cdot \bm{F}_{u,2,2} = \omega, \label{rot_av2}
\end{align}
for rotor 2. The positive and negative signs correspond to the same and opposite rotation directions, respectively.

Hereinafter, the effect of the concentration field of the surface-active compound released from the other rotor is treated as a perturbation. We first calculate the concentration field of the chemical compound released from one rotor and then the force originating from the concentration field exerting on the other particle.

For the construction of the concentration field generated by one camphor rotor, we consider that a single rotor is rotating at a constant angular velocity $\omega$, that is, the position is described as $\bm{r} = a \bm{e}(\phi_{1}) = a \bm{e}(\omega t + \phi_{0})$. We introduce a co-rotating frame with the rotor, where the variables in the frame are expressed with tildes. The single point source is located at $\tilde{\bm{r}} = a \tilde{\bm{e}}_x$ and the concentration field $\tilde{u}$ in the co-rotating frame should be in a steady state. Here, $\tilde{\bm{e}}_x$ is a unit vector in the $\tilde{x}$ direction in the co-rotating frame. 
Then, the steady-state concentration field should hold
\begin{align}
- \omega \frac{\partial \tilde{u}}{\partial \tilde{\theta}} = \tilde{\nabla}^2 \tilde{u} - \tilde{u} + \delta \left( \tilde{\bm{r}} - a\tilde{\bm{e}}_x \right),
\end{align}
where $\tilde \nabla$ is the nabla operator in the co-rotating frame.
After lengthy calculation, we obtain
\begin{align}
\tilde{u}(\tilde{r}, \tilde{\theta}) = \left\{ \begin{array}{ll} \tilde{u}_\mathrm{in}(\tilde{r}, \tilde{\theta}), & \tilde{r} < a, \\ \tilde{u}_\mathrm{out}(\tilde{r}, \tilde{\theta}), & \tilde{r} \geq a,
\end{array} \right.
\end{align}
where 
\begin{align}
& \tilde{u}_{\rm in}(\tilde{r}, \tilde{\theta}) \nonumber \\
& \qquad = \frac{1}{2\pi} \sum_{n=-\infty}^\infty \mathcal{K}_n  \left( a \sqrt{1 - i n \omega} \right) \mathcal{I}_n  \left( \tilde{r} \sqrt{1 - i n \omega} \right) e^{i n \tilde{\theta}}, \label{single_conc_in}
\end{align}
and
\begin{align}
&\tilde{u}_{\rm out}(\tilde{r}, \tilde{\theta}) \nonumber \\
& \qquad = \frac{1}{2\pi} \sum_{n=-\infty}^\infty \mathcal{I}_n  \left( a \sqrt{1 - i n \omega} \right) \mathcal{K}_n  \left(\tilde{r} \sqrt{1 - i n \omega} \right) e^{i n \tilde{\theta}}. \label{single_conc_out}
\end{align}
Here, $\mathcal{I}_n(\cdot)$ and $\mathcal{K}_n(\cdot)$ are the modified Bessel functions of the first and second kinds with the degree of $n$, respectively. The detailed derivation and notes for the Bessel function with complex parameters are shown in Appendix \ref{App_corotating}.

Based on Eq.~\eqref{single_conc_out}, we obtain the asymptotic form of the concentration field far from the rotor. Here we consider $\tilde{r} \gg 1 \gg a$. That is to say, we consider the case that the distance between two rotors is much greater than the diffusion length and the arm of the rotor is much less than the diffusion length. We take into account the order up to $\mathcal{O}(a)$. Using Eq.~\eqref{single_conc_out}, we obtain the asymptotic form as
\begin{align}
\tilde{u}(\tilde{r}, \tilde{\theta}) \simeq &  \frac{1}{2\sqrt{2\pi \tilde{r}}} e^{-\tilde{r}} + \frac{a}{2\sqrt{2\pi \tilde{r}}} \rho^{1/4} e^{ - \tilde{r} \sqrt{\rho} \cos(\chi/2)} \nonumber \\
&  \quad \times \cos \left(\tilde{\theta} - \frac{\chi}{4} + \tilde{r} \sqrt{\rho}\sin\frac{\chi}{2} \right) + \mathcal{O}\left( a^2\right) . \label{asymptotic_0}
\end{align}
Here, we set $1 - i \omega = \rho e^{-i \chi}$, that is, $\rho = \sqrt{1 + \omega^2}$ and $\chi = \arctan \omega$. The detailed calculation is found in Appendix~\ref{asymptotic}.

In the laboratory frame, the asymptotic form of the concentration field generated by a rotor, which is located at the origin, whose phase is $\phi$, and which rotates in the counterclockwise rotation, is described as
\begin{align}
& u(r, \theta, \phi) \nonumber \\
&\simeq \frac{1}{2\sqrt{2\pi r}} e^{-r} + \frac{a}{2\sqrt{2\pi r}} \rho^{1/4} e^{ - r \sqrt{\rho} \cos(\chi/2)} \nonumber \\
&  \qquad \times \cos \left( \theta - \phi - \frac{\chi}{4} + r \sqrt{\rho}\sin\frac{\chi}{2} \right) + \mathcal{O}\left( a^2\right), \label{eq_asymptotic}
\end{align}
for $r \gg 1$ and $t \to \infty$.
It should be noted that the first term with $\mathcal{O}(1)$ does not depend on time and that the second term with $\mathcal{O}(a)$ depends on time which can induce the synchronization between multiple rotors.

We adopt the asymptotic form in Eq.~\eqref{eq_asymptotic} for $u_j$ to calculate $\bm{F}_{u,i,j}$ $(i\neq j)$ and consider the interaction between two rotors in Eq.~\eqref{eq_Fij}. 
Considering that the asymptotic form of $u_j$ is described as a function of $\phi_j$ and that the position $\bm{r}_i$ is a function of $\phi_i$, the force $\bm{F}_{u,i,j}$ is a function of $\phi_i$ and $\phi_j$, i.e., $\bm{F}_{u,i,j}(\phi_i, \phi_j)$. We assume that the interaction is so weak that the phase difference hardly changes in one period $2\pi/\omega$ and we can adopt the averaging method in the phase description~\cite{Kuramoto}.

First, we calculate the time evolution of $\phi_1$ from Eqs.~\eqref{eq_rot1} and \eqref{rot_av1} as
\begin{align}
     \frac{d\phi_1}{dt} 
    =& \frac{1}{\eta A a} \bm{e}\left(\phi_1 + \frac{\pi}{2} \right) \cdot \bm{F}_{u,1} \nonumber \\
    =& \omega + \frac{1}{\eta A a} \bm{e}\left(\phi_1 + \frac{\pi}{2} \right) \cdot \bm{F}_{u,1,2}(\phi_1,\phi_2), \nonumber \\
    \simeq & \omega + \frac{\omega}{2\pi \eta A a} \int_0^{2\pi/\omega}  \bm{e}\left(\phi_1 + \frac{\pi}{2} \right) \cdot \bm{F}_{u,1,2}(\phi_1,\phi_2) dt \nonumber \\
   =& \omega + \frac{1}{2\pi \eta A a} \nonumber \\
    & \times \int_0^{2\pi} \bm{e}\left(\phi_1 + \frac{\pi}{2} \right) \cdot \bm{F}_{u,1,2} (\phi_1, \phi_1 + \Delta \phi) d\phi_1 \nonumber \\
    \equiv & \left\{ \begin{array}{ll} \omega + G_s(\Delta \phi), & (\mathrm{same \, rotation \, direction}), \\ \omega + G_o(\Delta \phi), & (\mathrm{opposite \, rotation \, direction}). \end{array} \right.     \label{aveq_rot1}
\end{align}
Here, we set $\Delta \phi = \phi_2 - \phi_1$ and calculate the integral under the assumption that $\Delta \phi$ is constant.
It should be noted that $G_s(\Delta\phi)$ and $G_o(\Delta\phi)$ are different since $\bm{F}_{u,1,2}$ depends on the rotation direction of rotor 2.
In the same manner, we obtain from Eqs.~\eqref{eq_rot2} and \eqref{rot_av2} as
\begin{align}
     \frac{d\phi_2}{dt} 
    \simeq & \omega + \frac{1}{2\pi \eta A a} \nonumber \\ &  \times \int_0^{2\pi} \bm{e}\left(\pm \left(\phi_2 - \frac{\pi}{2} \right)\right) \cdot \bm{F}_{u,2,1} (\phi_2 - \Delta \phi, \phi_2) d\phi_2 \nonumber \\
= & \left\{ \begin{array}{ll} \omega + G_s(-\Delta \phi), & (\mathrm{same \, rotation \, direction}), \\ \omega + G_o(-\Delta \phi), & (\mathrm{opposite \, rotation \, direction}), \end{array} \right.  \label{aveq_rot2}
\end{align}
where the positive and negative signs in the second term on the right side correspond to the cases with the same and opposite rotation directions, respectively. $G_s( \Delta \phi )$ and $G_o( \Delta \phi )$ are the so-called phase coupling functions in terms of coupled oscillators~\cite{RMP_synchronization}.

In the case with the same rotation direction, the second term on the right side in Eq.~\eqref{aveq_rot2} is calculated as
\begin{align}
G_s(\Delta \phi)
&\simeq \frac{\Gamma \rho^{1/4} e^{-\sqrt{\rho} L\cos(\chi/2) }}{4 \eta \sqrt{2\pi L}} \nonumber \\
&\quad \times \left\{ \frac{1}{2L} \sin \left[ \Delta \phi + \frac{\chi}{4} - L \sqrt{\rho} \sin \left( \frac{\chi}{2}\right) \right] \right. \nonumber \\
& \quad \left. \quad - \sqrt{\rho} \sin \left[\Delta \phi + \frac{3\chi}{4} - L \sqrt{\rho} \sin \left( \frac{\chi}{2}\right)\right] \right\} \nonumber \\
&\equiv g_s( \Delta \phi ), \label{phase_eq_phi2s}
\end{align}
which is plotted in Fig.~\ref{fig6}(a).
The detailed calculation is shown in Appendix~\ref{app:phase_reduction}. Then, we have
\begin{align}
\frac{d\phi_1}{dt} \simeq& \omega + g_s(\Delta \phi), \\
\frac{d\phi_2}{dt} \simeq& \omega + g_s(-\Delta \phi).
\end{align}
By calculating the difference between two equations, we obtain the time-evolution equations for the slow dynamics of $\Delta \phi$ as
\begin{align}
\frac{d \Delta\phi}{dt} 
\simeq& g_s(-\Delta \phi) - g_s(\Delta \phi)
\equiv h_s(\Delta \phi), \label{Hs}
\end{align}
where
\begin{align}
h_s(\Delta \phi) =&\frac{\Gamma \rho^{1/4} e^{-\sqrt{\rho} L\cos(\chi/2) }}{2\eta \sqrt{2\pi L}} \sin \Delta \phi \nonumber \\
& \times \left\{ -\frac{1}{2L} \cos\left[ -\frac{\chi}{4} + L \sqrt{\rho} \sin \left( \frac{\chi}{2} \right)\right] \right. \nonumber \\
& \left. \quad + \sqrt{\rho} \cos \left[ -\frac{3\chi}{4}  + L \sqrt{\rho} \sin \left(\frac{\chi}{2} \right) \right] \right\} \nonumber \\
\equiv& \frac{\Gamma \rho^{1/4} e^{-\sqrt{\rho} L\cos(\chi/2) }}{2\eta \sqrt{2\pi L}} C_s \sin \Delta \phi \nonumber \\
\equiv & h_s^{(1)} \sin \Delta \phi.
\label{eq_pr_s}
\end{align}
As shown in Fig.~\ref{fig6}(c), $\Delta \phi = 0$ and $\pi$ are fixed points of Eq.~\eqref{Hs} and their stability is determined by the sign of $C_s$ since the factors other than $C_s$ are positive. That is to say, the fixed point at $\Delta \phi = 0$ is stable when $C_s < 0$ and it is unstable when $C_s > 0$. On the other hand, the fixed point at $\Delta \phi = \pi$ is unstable when $C_s < 0$ and it is stable when $C_s > 0$. Thus, when $C_s < 0$ and $C_s > 0$, in-phase and anti-phase synchronization should occur, respectively. It should be noted that $C_s$ depends only on $\omega$ and $L$ since $\rho$ and $\chi$ are functions of $\omega$, as shown just below Eq.~\eqref{asymptotic_0}. $C_s$ and $h_s^{(1)}$ for $\eta=0.1$ are plotted as a function of $L$ in Fig.~\ref{fig7}, where $\omega$ is set to be constant at $13.23$ from the numerical results in Fig.~\ref{fig2}.

\begin{figure}
\includegraphics{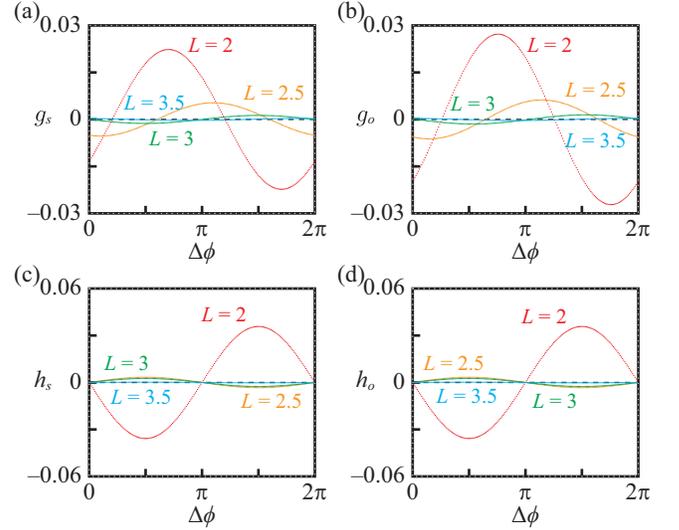}
\caption{\label{fig6} Plots of (a) $g_s$, (b) $g_o$, (c) $h_s$, and (d) $h_o$ against $\Delta \phi$ for $L = 2$, $2.5$, $3$, and $3.5$. We adopt $\omega = 13.23$.}
\end{figure}

\begin{figure}
\includegraphics{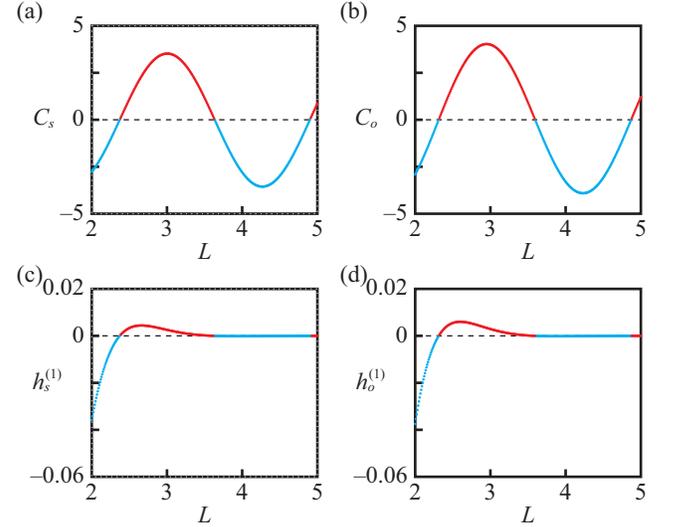}
\caption{\label{fig7} Plots of (a) $C_s$, (b) $C_o$, (c) $h_s^{(1)}$, and (d) $h_o^{(1)}$ against $L$. Here, we adopt $\omega = 13.23$. Positive (red) and negative (cyan) signs represent the preference of the anti-phase and in-phase synchronization, respectively.}
\end{figure}

In the case with the opposite rotation direction, we obtain in the same manner as in the case with the same rotation direction,
\begin{align}
G_o(\Delta\phi) 
& \simeq -\frac{\Gamma \rho^{1/4} e^{-\sqrt{\rho} L\cos(\chi/2) }}{4 \eta \sqrt{2\pi L}} \nonumber \\
&\quad \times \left\{ \frac{3}{2L} \sin \left[ \Delta \phi + \frac{\chi}{4} - L \sqrt{\rho} \sin \left( \frac{\chi}{2}\right) \right] \right. \nonumber \\
& \left. \quad + \sqrt{\rho} \sin \left[ \Delta \phi + \frac{3\chi}{4} - L \sqrt{\rho} \sin \left( \frac{\chi}{2}\right)\right] \right\} \nonumber \\
&\equiv g_o( \Delta \phi ), \label{phase_eq_phi2o}
\end{align}
which leads to
\begin{align}
\frac{d \Delta\phi}{dt} \simeq g_o(-\Delta\phi) - g_o(\Delta \phi) \equiv h_o(\Delta \phi). \label{Ho}
\end{align}
Here, we have calculated $h_o(\Delta \phi)$ as
\begin{align}
h_o(\Delta \phi) =& \frac{\Gamma \rho^{1/4} e^{-\sqrt{\rho} L\cos(\chi/2) }}{2\eta \sqrt{2\pi L}} \sin \Delta \phi \nonumber \\
& \times \left\{ \frac{3}{2L} \cos\left[ -\frac{\chi}{4} + L \sqrt{\rho} \sin \left( \frac{\chi}{2} \right)\right] \right. \nonumber \\
& \left. \quad + \sqrt{\rho} \cos \left[ -\frac{3\chi}{4}  + L \sqrt{\rho} \sin \left(\frac{\chi}{2} \right) \right] \right\} \nonumber \\
\equiv& \frac{\Gamma \rho^{1/4} e^{-\sqrt{\rho} L\cos(\chi/2) }}{2\eta \sqrt{2\pi L}} C_o \sin \Delta \phi \nonumber \\
\equiv & h_o^{(1)} \sin \Delta \phi. \label{eq_pr_o}
\end{align}
The plots of $g_o(\Delta\phi)$ and $h_o(\Delta \phi)$ are shown in Figs.~\ref{fig6}(b) and (d).
In this case, $\Delta \phi = 0$ and $\pi$ are also the fixed points. We can discuss the stability of the synchronization mode in the parallel manner, and thus the sign of the coefficient $C_o$ plays an important role. To exemplify the stable synchronization mode, $C_o$ are also plotted against $L$ in Fig.~\ref{fig7}. The signs of $C_s$ and $C_o$ almost coincide for fixed $L$, which means that the stable synchronization mode is the same in the cases with the same and opposite rotation directions for each $L$.

\section{Discussion \label{sec:discussion}}

\begin{figure}
\includegraphics{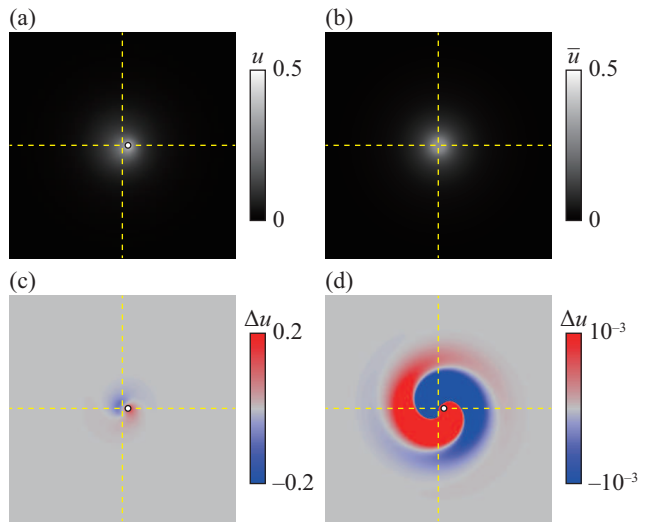}
\caption{\label{fig8} (a) Snapshot of the concentration field for a single rotor rotating at a constant angular velocity at $t = t_0 \simeq 100.1436$, when $\phi_1 = 0$ first holds after $t = 100$. (b) Averaged concentration field $\bar{u}$ over a period. (c) Difference $\Delta u$ between the concentration field $u$ in (a) from the averaged field $\bar{u}$ in (b). (d) Enhanced profile of (c), where the color range for concentration is magnified. The region with the size $8 \times 8$ is shown.}
\end{figure}

Here, we discuss the mechanism on the synchronization of the coupled rotors based on the theoretical results. The concentration field of the surface-active compound that one rotor releases is expressed in Eq.~\eqref{eq_asymptotic}. The first term on the right side does not depend on the phase, but the second term does.
The effect of the dynamics of the other rotor is approximately obtained by averaging the effect over one period. In such an averaging process, the effect of the first term is canceled out and only the second term matters. This means that the periodically changing concentration field only affects the stability of the synchronization mode. To visualize the time-dependent component of the concentration field, we numerically calculated the averaged concentration field $\bar{u}$,
\begin{align}
    \bar{u}(x,y) = \frac{1}{T} \int_{t_0}^{t_0 + T} u(x,y,t) dt,
\end{align}
where $t_0$ was the time when the rotor motion reached the stationary state and corresponded to $\phi_1 \simeq 0$.
Figure~\ref{fig8} shows the plot of $\Delta u(x,y,t_0)$, where $\Delta u(x,y,t) = u(x,y,t) - \bar{u}(x,y)$.
The time-dependent component $\Delta u$ has a spiral structure, which is consistent with the second term on the right side of Eq.~\eqref{eq_asymptotic}.
The pitch of the spiral $L_0$ is almost double of the length $L$ for which the stable synchronization mode changes. Considering that such spiral structure mainly comes from the second term on the right-hand side in Eq.~\eqref{eq_asymptotic}, the pitch $L_0$ is estimated from
\begin{align}
    L_0 \sqrt{\rho} \sin \frac{\chi}{2} = 2 \pi.  
\end{align}
Actually, $L_0$ is calculated as 2.54 with $\omega = 13.23$. This value for $L_0$ well corresponds to both the results by numerical simulations and theoretical analyses, where the stability of the synchronization mode changes every $\simeq 1.2$ in $L$.

\begin{figure}
\includegraphics{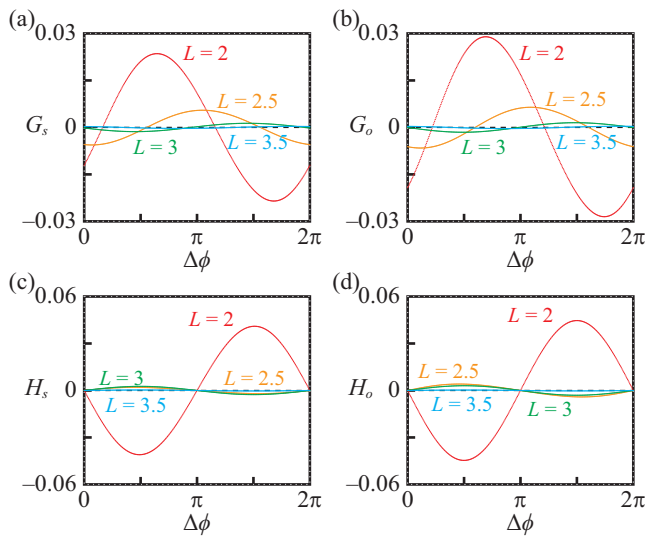}
\caption{\label{fig9} Plots of (a) $G_s$, (b) $G_o$, (c) $H_s$, and (d) $H_o$ against $\Delta \phi$. The results with $L = 2$ (red), $L = 2.5$ (orange), $L = 3$ (green), and $L = 3.5$ (cyan) are shown in each panel.  }
\end{figure}

\begin{figure}
\includegraphics{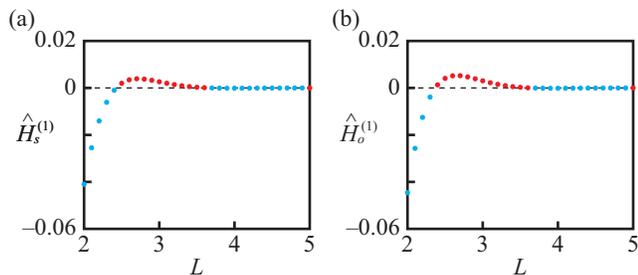}
\caption{\label{fig10} Plots of $\hat{H}_s^{(1)}$ and $\hat{H}_o^{(1)}$ against $L$. The positive and negative values are indicated with red and cyan points, respectively. The positive $\hat{H}_s^{(1)}$ or $\hat{H}_o^{(1)}$ means that the anti-phase synchronization state is stable, while the negative $\hat{H}_s^{(1)}$ or $\hat{H}_o^{(1)}$ means that the in-phase synchronization state is stable.}
\end{figure}

Next, we directly calculated the phase coupling function using the numerical simulation in order to justify the approximation adopted in the theoretical analysis. We directly calculated the functions $G_s(\Delta \phi)$ and $G_o (\Delta \phi)$ in Eqs.~\eqref{aveq_rot1} and \eqref{aveq_rot2}. $H_s(\Delta \phi)$ and $H_o (\Delta \phi)$ are defined as
\begin{align}
H_i(\Delta \phi) = G_i(-\Delta \phi) - G_i(\Delta \phi)
\end{align}
where $i$ denotes $s$ or $o$. The plots for $G_s(\Delta \phi)$, $G_o(\Delta\phi)$, $H_s(\Delta \phi)$, and $H_o(\Delta \phi)$ obtained from the numerical calculation are
shown in Fig.~\ref{fig9}. In the calculation, we first calculated the dynamics of the $i$th rotor only considering the concentration field released from itself until it reached a stationary angular velocity, and then calculated the force $\bm{F}_{u,j,i}$ working on the $j$th particle rotating at the given angular velocity, which was the same as the $i$th rotor's. Using the obtained force, $G_s(\Delta\phi)$ and $G_o(\Delta \phi)$ were calculated by averaging over a period.
The numerical simulation was performed in the same procedure as in Sec.~\ref{sec:modelling}. We detected $\tau^{(1)}_\nu$ just after $t = 100$ and calculated the average from $t = \tau^{(1)}_\nu$ to $\tau^{(1)}_{\nu + 1}$. $G_i$ and $H_i$ were calculated for $\Delta \phi = 2 \pi \lambda/ 360$ where $\lambda = 0, \cdots , 359$, and the averaging was performed for each time step during one period. The functions $H_s(\Delta \phi)$ and $H_o(\Delta \phi)$ are odd functions, which take the value of zero at $\Delta \phi = 0, \pi$ and have one positive and one negative peak. In order to discuss the stability in the synchronization mode, the slopes at $\Delta \phi = 0, \pi$ are important. Therefore, we consider the Fourier sine expansion of the functions as
\begin{align}
H_i(\Delta \phi) = \sum_{k=1}^\infty \hat{H}_i^{(k)} \sin k \Delta \phi.
\end{align}
where $i$ denotes $s$ or $o$. The first-mode coefficient determines the stability of the synchronization mode; the in-phase and anti-phase synchronization is stable for $\hat{H}_i^{(1)} < 0$ and $\hat{H}_i^{(1)} > 0$, respectively. In Fig.~\ref{fig10}, $\hat{H}_s^{(1)}$ and $\hat{H}_o^{(1)}$ are plotted against $L$, which is close to the plot in Figs.~\ref{fig7}(c) and (d) obtained by theoretical analysis.

In the calculation of the phase coupling function shown in the previous paragraph, we assume that the two rotors rotate at a constant angular velocity, i.e., intrinsic angular velocity; however, the angular velocity of the rotor should be affected by the concentration field originating from the other rotor. Therefore, we also calculated the phase coupling function including the effect of the change in the angular velocity, and found that the time evolution of the phase difference is qualitatively the same as the one shown in Figs.~\ref{fig9} and \ref{fig10}. The details are shown in Appendix~\ref {app:phase_response_function}.

Here, we discuss the parameter set used in the numerical simulation. From the previous reports on experiments~\cite{kitahata_rectangle}, the diffusion length is considered to be several tens of millimeters. Considering that the unity in the length scale is set to be the diffusion length, the arm length of the rotor and the distance between the rotors are estimated to be several millimeters and several centimeters, respectively. These scales are in the same scale adopted in the experimental setup~\cite{Sharma_PRE.103.012214}. The time scale is normalized by the characteristic evaporation time, which is around several seconds. The angular velocity of the rotors and the characteristic time for the relaxation time to the synchronization mode should be affected by the combination of the particle size, supply rate of the chemicals, and proportionality coefficient between the surface tension and concentration. Since these parameters are difficult to be directly measured from the experiments, the more precise correspondence to the experimental system should be done as a future study.

In the present system, the stable synchronization mode changes depending on the distance between the two rotors. There have been several studies of similar behaviors in the other systems such as the cell thickness pattern in slime mold~\cite{Takamatsu} and the coupled system of a flickering candle flame~\cite{Kitahata_candle,Manoj2018}. The time delay in the interaction plays an important role in the former case, while the nonlinear coupling manner is dominant in the latter case. In the coupling between the camphor rotor discussed in the present paper, the interaction between two rotors is through the concentration field that obeys the linear equation, and thus the time delay seems to play an important role in the present system. Actually, the spiral structure shown in Fig.~\ref{fig8} is the result of the supply from the rotating rotor and diffusion. Due to the linearity of the equation, we succeeded to write the time-delay effect directly through the concentration field, and such time-delay effect is reduced to the interaction term in the phase dynamics.

\section{Conclusion \label{sec:conclusion}}

We investigated the coupled active rotors, which spontaneously exhibit rotation due to the surface tension gradient originating from the surface-active chemicals released from itself. It has been reported that by experiments such a coupled rotor system show both in-phase and anti-phase synchronization, but the mechanism was not fully clarified. We consider the mathematical model which includes the time evolution of the concentration field and the motion of the rotors, and obtain the results that the stable synchronization mode alternates between in-phase and anti-phase synchronization with an increase in the distance between the two rotors. By adopting the phase description, which has often been used for the coupled oscillator systems, we derive the time evolution equation for the phase difference between the two rotors. The theoretical results suggest the alternate stable synchronization mode depending on $L$, which well corresponds to the numerical calculation results. We also directly evaluated the phase coupling function from the numerical calculation and confirmed that our theoretical approach works well. We hope that our results will be reproduced in the experimental systems with two active rotors with well-controlled intrinsic angular velocities. As for the extension of the present study, three-or-more rotor systems should be interesting since the system has many possible stable modes and may exhibit chaotic behaviors. As another extension, the theoretical description for the case with strong interaction should be interesting. In such a case, the rotation of each rotor should change to the reciprocal motion along an arc and such reciprocal motion of the two rotors can synchronize.

\begin{acknowledgments}
This work was supported by JSPS KAKENHI Grants No.~JP19K03765, No.~JP20K14370, No.~JP20H02712, and No.~JP21H01004, and also the Cooperative Research Program of ``Network Joint Research Center for Materials and Devices'' (Grants No.~20224003 and No.~20221173).
This work was also supported by JSPS and PAN under the Japan-Poland Research Cooperative Program (Grant No.~JPJSBP120204602).
\end{acknowledgments}

\appendix

\section{Derivation of concentration field in a co-rotating frame with a single rotor\label{App_corotating}}

In this section, we derive a steady concentration field in a co-rotating frame with a single rotor with an angular velocity $\omega$.
The positions in the original system and in the co-rotating system are denoted as $\bm{r} = {}^{t} (r \cos\theta, r \sin\theta)$ and $\tilde{\bm{r}} = {}^t (\tilde{r} \cos \tilde{\theta}, \tilde{r} \sin \tilde{\theta})$, respectively. Then, we have
\begin{align}
\tilde{\bm{r}} = \mathcal{R}(-\omega t) {\bm{r}},
\end{align}
where $\mathcal{R}(\psi)$ is a matrix for rotation in a two-dimensional system, i.e.,
\begin{align}
\mathcal{R}(\psi) = \left( \begin{array}{cc} \cos\psi & -\sin\psi \\ \sin\psi & \cos\psi \end{array} \right).
\end{align}
In other words, we have the relation as
\begin{align}
\tilde{\theta} = \theta - \omega t.
\end{align}
Then the operator for the time derivative is rewritten in the $\tilde{\bm{r}}$ system as
\begin{align}
& \frac{\partial}{\partial t} + \omega \left( \begin{array}{cc} - \sin \omega t & \cos\omega t \\ -\cos\omega t & -\sin\omega t \end{array} \right) \bm{r} \cdot \tilde{\nabla} \nonumber \\
=& \frac{\partial}{\partial t} + r \omega \left( \begin{array}{cc} \sin (\theta - \omega t) \\ -\cos (\theta - \omega t) \end{array} \right)  \cdot \tilde{\nabla} \nonumber \\
=& \frac{\partial}{\partial t} + \tilde{r} \omega \left( \begin{array}{cc} \sin \tilde{\theta} \\ -\cos \tilde{\theta} \end{array} \right) \cdot \tilde{\nabla} \nonumber \\
=& \frac{\partial}{\partial t} - \omega \tilde{r} \tilde{\bm{e}}_\theta \cdot \tilde{\nabla}. 
\end{align}
Here $\tilde{\nabla}$ is the nabla operator in the $\tilde{\bm{r}}$ system and $\tilde{\bm{e}}_\theta$ is a unit vector in the co-rotating system, $\tilde{\bm{e}}_\theta = {}^t ( - \sin \tilde{\theta}, \cos \tilde{\theta})$.

The position of the point source $\tilde{\bm{a}}$ in the co-rotating frame is no longer dependent on time and is written as
\begin{align}
\tilde{\bm{a}} = a \tilde{\bm{e}}_x,
\end{align}
where $\tilde{\bm{e}}_x$ is a unit vector along the $\tilde{x}$ axis in the $\tilde{\bm{r}}$ system.

Hereafter, we omit the tilde ($\,\tilde{}\,$) for the variables in the co-moving frame. In order to consider the steady-state concentration field $u(r, \theta)$ in the co-rotating system, we set the time derivative to be 0. Therefore, the equation to be considered is explicitly written as
\begin{align}
- \omega \frac{\partial u}{\partial \theta} = \nabla^2 u - u + \delta \left(\bm{r} - \bm{a} \right). \label{time-evolution2}
\end{align}

The homogeneous equation for \eqref{time-evolution2} is expressed as
\begin{align}
- \omega \frac{\partial u}{\partial \theta} = \nabla^2 u - u. \label{homogeneous}
\end{align}
By assuming that Eq.~\eqref{homogeneous} has a solution in the form of
\begin{align}
u(r, \theta) = f_n(r) e^{i n \theta},
\end{align}
we obtain
\begin{align}
\frac{d^2f_n}{dr^2} + \frac{1}{r} \frac{df_n}{dr} - \frac{n^2}{r^2}f_n - (1 - in\omega) f_n = 0.  \label{complex_bessel}
\end{align}

By setting $\hat{r} = r \sqrt{1 - in\omega}$, we get 
\begin{align}
\frac{d^2f_n}{d {\hat{r}}^2} + \frac{1}{\hat{r}} \frac{df_n}{d \hat{r}} - \frac{n^2}{{\hat{r}}^2}f_n - f_n = 0,
\end{align}
which is the so-called modified Bessel equation and the solution is given as
\begin{align}
f_n(\hat{r}) = A_n \mathcal{I}_n \left( \hat{r} \right) + B_n \mathcal{K}_n  \left( \hat{r} \right).
\end{align}
Thus, $f_n(r)$ is described as
\begin{align}
f_n(r) =& A_n \mathcal{I}_n \left( r \sqrt{1 - in\omega} \right) + B_n \mathcal{K}_n \left(r \sqrt{1 - in\omega} \right) .
\end{align}
Then, the general solution of Eq.~\eqref{homogeneous} is given as
\begin{align}
u(r, \theta) =& \sum_{n=-\infty}^\infty \left[ A_n \mathcal{I}_n \left( r \sqrt{1 - i n \omega} \right) \right. \nonumber \\
&\left. \qquad + B_n \mathcal{K}_n  \left( r \sqrt{1 - i n \omega} \right) \right] e^{i n \theta}, \label{gensol}
\end{align}
where $A_n$ and $B_n$ are complex constants. Considering that $u$ is real, $A_n = A^*_{-n}$ and $B_n = B^*_{-n}$ should hold, where the superscript ``${}^*$'' indicates the complex conjugate.

It is notable that the solution of Eq.~\eqref{complex_bessel} should be complex. If we set it to be
\begin{align}
f_n(r) = p_n(r) + i q_n(r),
\end{align}
then we obtain
\begin{align}
\frac{d^2p_n}{dr^2} + \frac{1}{r} \frac{dp_n}{dr} - \frac{n^2}{r^2}p_n - p_n - n \omega q_n  = 0,
\end{align}
\begin{align}
\frac{d^2q_n}{dr^2} + \frac{1}{r} \frac{dq_n}{dr} - \frac{n^2}{r^2}q_n - q_n + n \omega p_n  = 0.
\end{align}
Considering that the modified Bessel function of the first kind, $\mathcal{I}_n(\cdot)$, is analytic in $\mathbb{C}$, and the modified Bessel function of the second kind, $\mathcal{K}_n(\cdot)$, is analytic in $\mathbb{C}$ except negative real numbers, the following relations hold:
\begin{align}
\mathcal{I}_n\left(z^*\right) = \left[ \mathcal{I}_n(z) \right]^*, \label{conjugate1}
\end{align}
\begin{align}
\mathcal{K}_n\left(z^*\right) = \left[ \mathcal{K}_n(z)\right]^*. \label{conjugate2}
\end{align}

These expressions are derived from Eq.~\eqref{gensol} considering that $\mathcal{I}_n(z)$ does not diverge at $\left| z \right| \to 0$ and that $\mathcal{K}_n(z)$ does not diverge at $\left| z \right| \to \infty$ for $\Re (z) > 0$. Now we set
\begin{align}
1 - i n \omega = \rho e^{-i \chi_n},
\end{align}
where $\rho > 0$ and $-\pi/2 < \chi_n < \pi/2$ holds, and define 
\begin{align}
\sqrt{1 - i n \omega} = \sqrt{\rho} e^{-i \chi_n / 2}. \label{sqrt1-inw}
\end{align}
Then, we can show that $\mathcal{K}_n \left( r \sqrt{1 - i n \omega} \right)$ does not diverge for $r \to \infty$. 

Thus, we can describe
\begin{align}
u_{\rm in}(r, \theta) =& A_0 \mathcal{I}_0(r) \nonumber \\
&+ 2\sum_{n=1}^\infty \Re \left[A_n \mathcal{I}_n \left( r \sqrt{1 - i n \omega} \right) e^{i n \theta} \right],
\end{align}
for $r < a$, and
\begin{align}
u_{\rm out}(r, \theta) =& B_0 \mathcal{K}_0(r) \nonumber \\
&+ 2 \sum_{n=1}^\infty \Re \left[B_n \mathcal{K}_n  \left( r \sqrt{1 - i n \omega} \right) e^{i n \theta} \right],
\end{align}
for $r > a$.

The coefficients $A_n$ and $B_n$ are determined by the condition that $u$ and $\nabla u$ are continuous at $r=a$ and that $u$ satisfies the inhomogeneous equation in Eq.~\eqref{time-evolution2}. From the continuity condition for $u$, we obtain
\begin{align}
 A_n\mathcal{I}_n \left(a \sqrt{1 - i n \omega} \right) = B_n \mathcal{K}_n \left(a \sqrt{1 - i n \omega} \right),
\end{align}
and thus we newly set $C_n$ so it holds that
\begin{align}
A_n = C_n \mathcal{K}_n \left(a \sqrt{1 - i n \omega}\right),
\end{align}
and
\begin{align}
B_n = C_n \mathcal{I}_n \left(a \sqrt{1 - i n \omega} \right).
\end{align}
Then, we calculate the difference in the derivative in the $r$ direction. Note that
\begin{align}
& \lim_{r \to a - 0} \frac{\partial u_{\rm in}}{\partial r} = \hat{C}_0 \mathcal{K}_0\left(\hat{a}_0 \right) \mathcal{I}_1\left(\hat{a}_0 \right) \nonumber \\
& \qquad + 2\sum_{n=1}^\infty \Re \left[ \hat{C}_n \mathcal{K}_n \left(\hat{a}_n \right) \frac{ \mathcal{I}_{n-1}\left(\hat{a}_n \right) + \mathcal{I}_{n+1}\left(\hat{a}_n \right) }{2} e^{i n \theta} \right],
\end{align} 
and
\begin{align}
& \lim_{r \to a + 0} \frac{\partial u_{\rm out}}{\partial r} = -\hat{C}_0 \mathcal{I}_0\left(\hat{a}_n \right) \mathcal{K}_1\left( \hat{a}_n \right) \nonumber \\
& \qquad - 2\sum_{n=1}^\infty \Re \left[ \hat{C}_n  \mathcal{I}_n \left(\hat{a}_n\right) \frac{ \mathcal{K}_{n-1}\left(\hat{a}_n \right) + \mathcal{K}_{n+1} \left(\hat{a}_n \right)}{2} e^{i n \theta} \right],
\end{align}
where $\hat{a}_n = a \sqrt{1 - i n \omega}$ and $\hat{C}_n = C_n \sqrt{1 - i n \omega}$.
Considering that $\hat{C}_0 = C_0$ and $\hat{a}_0 = a$, we obtain
\begin{align}
& \lim_{r \to a - 0} \frac{\partial u_{\rm in}}{\partial r} - \lim_{r \to a + 0} \frac{\partial u_{\rm out}}{\partial r} \nonumber \\
& =
C_0 \left[ \mathcal{K}_0\left( a \right) \mathcal{I}_1\left(a \right) + \mathcal{K}_1\left(a \right) \mathcal{I}_0\left(a \right) \right] \nonumber \\
& \quad +2 \sum_{n=1}^\infty \Re \biggl[ \hat{C}_n \left[ \mathcal{K}_n\left(\hat{a}_n \right) \mathcal{I}_{n+1}\left(\hat{a}_n \right) + \mathcal{K}_{n+1}\left(\hat{a}_n \right) \mathcal{I}_n\left(\hat{a}_n \right)  \right. \nonumber \\
& \qquad  \left.+ \mathcal{K}_n\left(\hat{a}_n \right) \mathcal{I}_{n-1}\left(\hat{a}_n \right) + \mathcal{K}_{n-1}\left(\hat{a}_n \right) \mathcal{I}_n\left(\hat{a}_n \right) \right] e^{in\theta}\biggr] \nonumber \\
& \quad = \frac{C_0}{a} +2\sum_{n=1}^\infty \Re \left[ \frac{\hat{C}_n}{\hat{a}_n} e^{i n \theta} \right] \nonumber \\
& \quad = \frac{C_0}{a} +2\sum_{n=1}^\infty \Re \left[ \frac{C_n}{a} e^{i n \theta} \right].
\end{align}
Therefore, we set $C_n = 1 / (2\pi)$ for all $n$, and we obtain
\begin{align}
\lim_{r \to a - 0} \frac{\partial u_{\rm in}}{\partial r} - \lim_{r \to a + 0} \frac{\partial u_{\rm out}}{\partial r} 
=& \frac{1}{2 \pi a} \sum_{n = -\infty}^\infty e^{i n\theta} \nonumber \\
=& \frac{1}{a} \delta (\theta),
\end{align}
which corresponds to the considered situation.

Considering that
\begin{align}
\int_0^{2\pi} \mathcal{I}_n \left( r \sqrt{1 - i n \omega} \right) e^{i n \theta} d\theta = 0 ,
\end{align}
\begin{align}
\int_0^{2\pi} \mathcal{K}_n \left( r \sqrt{1 - i n \omega} \right) e^{i n \theta} d\theta = 0 ,
\end{align}
for any nonzero integer $n$, and that
\begin{align}
& \int \left[ \omega \frac{\partial u}{\partial \theta} + \nabla^2 u - u + \delta \left( \bm{r} - a \bm{e}_x \right)  \right] d\bm{r} \nonumber \\
& \quad = -\int u( \bm{r}) d\bm{r} + 1 \nonumber \\
& \quad = 0, \label{integration}
\end{align}
we can explicitly execute the integration of $u(\bm{r})$ as
\begin{align}
& \int_0^{2\pi}\int_0^\infty u(r, \theta) d \bm{r} \nonumber \\
& \qquad = 2 \pi \left[ \int_0^a \frac{1}{2\pi} \mathcal{K}_0(a) \mathcal{I}_0(r) r dr \right. \nonumber \\ & \qquad \qquad \left. + \int_a^\infty \frac{1}{2\pi} \mathcal{I}_0(a) \mathcal{K}_0(r) r dr  \right] \nonumber \\
& \qquad  =  2\pi \left[\frac{1}{2\pi} \mathcal{K}_0(a) a \mathcal{I}_1(a) + \frac{1}{2\pi} \mathcal{I}_0(a) a \mathcal{K}_1(a) \right] \nonumber \\
& \qquad = a \left[\mathcal{K}_0(a) \mathcal{I}_1(a) + \mathcal{I}_0(a) \mathcal{K}_1(a) \right] \nonumber \\
& \qquad = 1,
\end{align}
which satisfies Eq.~\eqref{integration}.
 
Therefore, the steady-state concentration field for a single point source at $\tilde{\bm{r}} = a \tilde{\bm{e}}_x$ in the co-rotating frame is written as
\begin{align}
& u_{\rm in}(r, \theta) \nonumber \\
& = \frac{1}{2\pi} \mathcal{K}_0(a)\mathcal{I}_0(r) \nonumber \\
& \qquad + \sum_{n=1}^\infty \Re \left[\frac{1}{\pi} \mathcal{K}_n  \left( a \sqrt{1 - i n \omega} \right) \mathcal{I}_n  \left( r \sqrt{1 - i n \omega} \right) e^{i n \theta} \right] \nonumber \\
& = \frac{1}{2\pi} \sum_{n=-\infty}^\infty \mathcal{K}_n  \left( a \sqrt{1 - i n \omega} \right) \mathcal{I}_n  \left( r \sqrt{1 - i n \omega} \right) e^{i n \theta}, \label{single_conc_in_ap}
\end{align}
for $r < a$, and
\begin{align}
& u_{\rm out}(r, \theta) \nonumber \\
& = \frac{1}{2\pi}\mathcal{I}_0(a) \mathcal{K}_0(r) \nonumber \\
& \qquad + \sum_{n=1}^\infty \Re \left[\frac{1}{\pi} \mathcal{I}_n  \left( a \sqrt{1 - i n \omega} \right) \mathcal{K}_n  \left(r \sqrt{1 - i n \omega} \right) e^{i n \theta} \right] \nonumber \\
& = \frac{1}{2\pi} \sum_{n=-\infty}^\infty \mathcal{I}_n  \left( a \sqrt{1 - i n \omega} \right) \mathcal{K}_n  \left(r \sqrt{1 - i n \omega} \right) e^{i n \theta}, \label{single_conc_out_ap}
\end{align}
for $r > a$, which are Eqs.~\eqref{single_conc_in} and \eqref{single_conc_out} in the main text. In the calculation, we used the equalities given in Ref.~\cite{Bessel}.

\section{Derivation of the asymptotic form \label{asymptotic}}

Here, we derive the asymptotic form of the concentration field far from the source ($r \gg 1$) for $a \ll 1$. 

Considering the Maclaurin expansion of $\mathcal{I}_n(z)$ is given as~\cite{Bessel}
\begin{align}
\mathcal{I}_n(z) = \sum_{k=0}^\infty \frac{1}{k! \Gamma(n + k + 1)} \left(\frac{z}{2}\right)^{n + 2k}, \label{expansion_I}
\end{align}
the leading term of $I_n(z)$ is
\begin{align}
\mathcal{I}_n(z) = \frac{1}{2^n n!} z^n.
\end{align}
Therefore, we only need to consider $n=0$ and $n = \pm 1$ as far as we consider the first order of $a$.
Consider that the asymptotic form of $\mathcal{K}_n(z)$ is
\begin{align}
\mathcal{K}_n(z) \sim \sqrt{\frac{\pi}{2z}} e^{-z} \sum_{k = 0}^\infty \frac{\Gamma(n+k+1/2)}{k! \Gamma(n-k+1/2)} \frac{1}{(2z)^k},  \label{Kn_asymp}
\end{align}
for $\left|\arg z\right| < 3\pi/2$~\cite{Bessel}. Here, $\Gamma(\cdot)$ is the gamma function.
For $n = 0$ and $n = 1$, we obtain
\begin{align}
\mathcal{K}_0(z) = \sqrt{\frac{\pi}{2z}}e^{-z} \left( 1- \frac{1}{8z} + \mathcal{O}\left( \frac{1}{z^2} \right) \right), \label{K0}
\end{align}
\begin{align}
\mathcal{K}_1(z) = \mathcal{K}_{-1}(z) = \sqrt{\frac{\pi}{2z}}e^{-z} \left( 1+ \frac{3}{8z} + \mathcal{O}\left( \frac{1}{z^2} \right) \right), \label{K1}
\end{align}
Then, we only need to consider the terms with $n = 0, \pm 1$. By setting $\chi = \chi_1 = -\chi_{-1}$, we obtain
\begin{align}
&u_{\rm out}(r, \theta) \nonumber \\
&\quad = \frac{1}{2\pi} \left[ \mathcal{K}_0 \left(r \right) + \frac{a}{2} \sqrt{\rho} e^{-i\chi/2}  \mathcal{K}_1 \left(r \sqrt{\rho} e^{-i \chi/2} \right) e^{i \theta} \right. \nonumber \\
& \qquad \left. + \frac{a}{2} \sqrt{\rho} e^{i \chi/2}  \mathcal{K}_1 \left(r \sqrt{\rho} e^{i\chi/2} \right) e^{-i  \theta} \right] + \mathcal{O}(a^2) \nonumber \\
& \quad \simeq \frac{1}{2\pi} \left[ \sqrt{\frac{\pi}{2 r}} e^{-r} \right. \\
&\qquad \left.+ \frac{a}{2} \sqrt{\rho} e^{-i\chi/2}\frac{ \sqrt{\pi} }{ \sqrt{2r} \rho^{1/4} e^{-i\chi/4} }e^{-r \sqrt{\rho} e^{-i\chi/2}} e^{i\theta} \right. \nonumber \\
& \qquad \left.+ \frac{a}{2} \sqrt{\rho} e^{i\chi/2}\frac{ \sqrt{\pi} }{ \sqrt{2r} \rho^{1/4} e^{i\chi/4} }e^{-r \sqrt{\rho} e^{i\chi/2}} e^{-i\theta} \right] + \mathcal{O}\left(a^2\right) \nonumber \\
& \quad = \frac{1}{2 \sqrt{2 \pi r}} e^{-r}\left[ 1 + \frac{a}{2} \rho^{1/4} e^{r(1- \sqrt{\rho} \cos (\chi/2))} \right. \nonumber \\
& \qquad \times \left. \left( e^{i  (\theta - \chi/4 + r \sqrt{\rho} \sin (\chi/2) )} + e^{-i  (\theta - \chi/4 + r \sqrt{\rho} \sin (\chi/2) )} \right) \right] \nonumber \\
& \qquad  + \mathcal{O}\left( a^2 \right)\nonumber \\
& \quad = \frac{1}{2\sqrt{2\pi r}} e^{-r} + \frac{a}{2\sqrt{2\pi r}} \rho^{1/4} e^{ - r \sqrt{\rho} \cos(\chi/2)} \nonumber \\
&  \qquad \times \cos \left( \theta - \frac{\chi}{4} + r \sqrt{\rho}\sin\frac{\chi}{2} \right) + \mathcal{O}\left( a^2\right) ,
\end{align}
and thus we obtain Eq.~\eqref{asymptotic_0} in the main text.
Here, we only considered the leading terms in Eqs.~\eqref{K0} and \eqref{K1}, and used Eq.~\eqref{sqrt1-inw}. In the case that the phase of the rotor is $\phi_1$,  we obtain Eq.~\eqref{eq_asymptotic} by replacing $\theta$ with $\theta - \phi_1$.

\section{Calculation on the phase description \label{app:phase_reduction}}

Here, we show the detailed calculation of the time evolution using the averaging method.
From Eqs.~\eqref{eq_Fij} and \eqref{aveq_rot2}, we need to obtain $\left. - \nabla u \cdot \bm{e}(\phi_2 - \pi/2) \right|_{\bm{r} = \bm{r}_2}$.
Thus, we first calculate the gradient of the concentration field in the polar coordinates,
\begin{align}
    \nabla u = \frac{\partial u}{\partial r} \bm{e}(\theta) + \frac{1}{r} \frac{\partial u}{\partial \theta} \bm{e}\left( \theta + \frac{\pi}{2}\right).
\end{align}
The asymptotic form in Eq.~\eqref{eq_asymptotic} is separated into two parts,
\begin{align}
    u(r,\theta,\phi_1) = u^{(0)}(r) + u^{(1)}(r,\theta, \phi_1) a + \mathcal{O}(a^2), \label{expansion}
\end{align}
where
\begin{align}
    u^{(0)}(r) = \frac{1}{2 \sqrt{2\pi r}} e^{-r},
\end{align}
and
\begin{align}
    u^{(1)}(r,\theta,\phi_1) =& \frac{1}{2\sqrt{2\pi r}} \rho^{1/4} e^{ - r \sqrt{\rho} \cos(\chi/2)} \nonumber \\
&  \times \cos \left( \theta - \phi_1 - \frac{\chi}{4} + r \sqrt{\rho}\sin\frac{\chi}{2} \right).
\end{align}

Considering that $u^{(0)}$ does not depend on $\phi_1$ or $\phi_2$, $- \left. \nabla u^{(0)} \cdot \bm{e}(\phi_2 - \pi/2) \right|_{\bm{r} = \bm{r}_2}$ is a function of only $\phi_2$ and not $\phi_1$. As a result of the averaging, the dependence on $\phi_2$ should be omitted, and it gives only a constant value. Therefore, $u^{(0)}$ only secondarily affects the stability of the synchronization mode.

Therefore, we consider the effect by $u^{(1)}$. First we calculate the gradient of $u^{(1)}$ as
\begin{align}
    \nabla u^{(1)} =& \left[ -\frac{1}{2r} u^{(1)} - \sqrt{\rho} u^{(1)} \cos \frac{\chi}{2} - \sqrt{\rho} \hat{u}^{(1)} \sin \frac{\chi}{2} \right] \bm{e}(\theta) \nonumber \\
    & - \frac{1}{r}\hat{u}^{(1)} \bm{e}\left(\theta + \frac{\pi}{2}\right),
\end{align}
where we set
\begin{align}
    \hat{u}^{(1)} =& \frac{1}{2\sqrt{2\pi r}}\rho^{1/4} e^{-\sqrt{\rho} \cos(\chi/2) r} \nonumber \\
    & \quad \times \sin \left[\theta-\phi_1-\frac{\chi}{4} + \sqrt{\rho} \sin  \left(\frac{\chi}{2} \right) r\right].
\end{align}

Hereafter, we separately calculate the two cases, i.e., the case with the same rotation direction and that with the opposite rotation direction. First, we consider the case with the same rotation.
We calculate $\nabla u_1 \cdot \bm{e}(\phi_2 - \pi/2)$ at $\bm{r} = \bm{r}_2 = L \bm{e}_x - a \bm{e}(\phi_2)$. Considering that $r=\sqrt{L^2 + a^2 - 2La \cos\phi_2}$ and $\tan \theta = - a \sin \phi_2 / (L-a \cos \phi_2)$ in the polar coordinates, we obtain 
\begin{align}
    &\left. -\nabla u^{(1)} \cdot \bm{e}\left(\phi_2 - \frac{\pi}{2}\right) \right|_{\bm{r} = \bm{r}_2} \nonumber \\
    & \quad= \left. -\frac{\partial u^{(1)}}{\partial r} \bm{e}(\theta) \cdot \bm{e}\left(\phi_2 - \frac{\pi}{2} \right) \right|_{\bm{r} = \bm{r}_2} \nonumber \\
    & \qquad \left. - \frac{1}{r} \frac{\partial u^{(1)}}{\partial \theta} \bm{e}\left(\theta + \frac{\pi}{2}\right) \cdot \bm{e}\left(\phi_2 - \frac{\pi}{2} \right)\right|_{\bm{r} = \bm{r}_2} \nonumber \\
    &\quad = \left. \left( \frac{1}{2r} + \sqrt{\rho} \cos \frac{\chi}{2} \right) u^{(1)} \sin(\phi_2-\theta)\right|_{\bm{r} = \bm{r}_2} \nonumber \\
    &\qquad \left. - \left(\frac{1}{r} \cos(\phi_2-\theta) - \sqrt{\rho} \sin \frac{\chi}{2} \sin(\phi_2-\theta) \right) \hat{u}^{(1)} \right|_{\bm{r} = \bm{r}_2}. \label{force_order1_s}
\end{align}
By considering the Maclaurin expansion of $r$ and $\theta$ with respect to $a$, we obtain
\begin{align}
    r =& L + \mathcal{O}(a). \label{expand_r}
\end{align}
and
\begin{align}
    \theta = \mathcal{O}(a).
\end{align}
Therefore, we obtain
\begin{align}
    \sin(\phi_2-\theta) =& \sin \phi_2 + \mathcal{O}(a), \label{sinphi2theta}
\end{align}
\begin{align}
    \cos(\phi_2-\theta) =& \cos \phi_2 + \mathcal{O}(a). \label{eq_cos}
\end{align}
We also calculate $u^{(1)}$ and $\hat{u}^{(1)}$ at $\bm{r} = \bm{r}_2$ as
\begin{align}
& \left. u^{(1)} \right|_{\bm{r} = \bm{r}_2} \nonumber \\
    & = \frac{\rho^{1/4} e^{-\sqrt{\rho} (L-a\cos\phi)\cos(\chi/2)}}{2\sqrt{2\pi (L - a \cos \phi_2)}}  \cos \left[- \frac{a}{L} \sin\phi_2 -\phi_1- \frac{\chi}{4} \right. \nonumber \\
    & \qquad \left.+ \sqrt{\rho}(L-a\cos\phi_2) \sin \left(\frac{\chi}{2}\right)  \right] \nonumber \\
    &= \frac{\rho^{1/4}  e^{-\sqrt{\rho} L \cos(\chi/2)}}{2 \sqrt{2\pi L}} \cos \left[ -\phi_1- \frac{\chi}{4} + \sqrt{\rho} \sin \left(\frac{\chi}{2}\right) L \right] \nonumber \\
    & \quad + \mathcal{O}(a), \label{u_s}
\end{align}
and 
\begin{align}
    &\left. \hat{u}^{(1)}\right|_{\bm{r} = \bm{r}_2} \nonumber \\
    &= \frac{ \rho^{1/4} e^{-\sqrt{\rho} (L-a\cos\phi_2)\cos(\chi/2)}}{2\sqrt{2\pi (L-a\cos\phi) }} \sin \left[- \frac{a}{L} \sin\phi_2 -\phi_1  - \frac{\chi}{4} \right. \nonumber
    \\
    & \qquad \left. + \sqrt{\rho} (L-a\cos\phi_2) \sin \left(\frac{\chi}{2}\right) \right] \nonumber \\
    &= \frac{\rho^{1/4} e^{-\sqrt{\rho} L \cos(\chi/2)}}{2\sqrt{2\pi L}} \sin \left[ -\phi_1- \frac{\chi}{4} + \sqrt{\rho} \sin \left(\frac{\chi}{2}\right) L \right] \nonumber \\
    & \quad + \mathcal{O}(a).  \label{utilde_s}
\end{align}

Equation~\eqref{force_order1_s} with Eqs.~\eqref{sinphi2theta}--\eqref{utilde_s} leads to
\begin{align}
&\left. - \nabla u^{(1)} \cdot \bm{e}\left(\phi_2 - \frac{\pi}{2} \right)\right|_{\bm{r} = \bm{r}_2} \nonumber \\
&\quad = \frac{\rho^{1/4} e^{-\sqrt{\rho} L\cos(\chi/2) }}{2\sqrt{2\pi L}} \nonumber \\
& \qquad \times \left\{
-\frac{1}{2L} \sin \left[-\phi_2 -\phi_1- \frac{\chi}{4} + L \sqrt{\rho} \sin \left(\frac{\chi}{2}\right)  \right] \right. \nonumber \\
& \qquad \left. - \frac{1}{2L} \cos\phi_2 \sin \left[ -\phi_1- \frac{\chi}{4} + L\sqrt{\rho} \sin \left(\frac{\chi}{2}\right)  \right] \right. \nonumber \\
& \qquad \left. + \sqrt{\rho} \sin \phi_2 \cos \left[ -\phi_1- \frac{3\chi}{4} + L \sqrt{\rho} \sin \left(\frac{\chi}{2}\right) \right] \right\} \nonumber \\
& \qquad + \mathcal{O}(a),
\end{align}
and thus Eqs.~\eqref{aveq_rot2} and \eqref{expansion} give
\begin{align}
& \frac{d\phi_2}{dt} \simeq \omega + \frac{\Gamma \rho^{1/4} e^{-\sqrt{\rho} L\cos(\chi/2) }}{4 \eta \sqrt{2\pi L}} \nonumber \\
& \times \left\{ - \frac{1}{2L} \sin \left[ \Delta \phi - \frac{\chi}{4} + L \sqrt{\rho} \sin \left( \frac{\chi}{2}\right) \right] \right. \nonumber \\
& \left. \quad + \sqrt{\rho} \sin \left[\Delta \phi - \frac{3\chi}{4} + L \sqrt{\rho} \sin \left( \frac{\chi}{2}\right)\right] \right\}.
\end{align}
Considering geometric symmetry, we obtain
\begin{align}
& \frac{d\phi_1}{dt} \simeq \omega + \frac{\Gamma \rho^{1/4} e^{-\sqrt{\rho} L\cos(\chi/2) }}{4 \eta \sqrt{2\pi L}} \nonumber \\
& \times \left\{ -\frac{1}{2L} \sin \left[-\Delta \phi - \frac{\chi}{4} + L \sqrt{\rho} \sin \left( \frac{\chi}{2}\right) \right] \right. \nonumber \\
& \left. \quad + \sqrt{\rho} \sin \left[-\Delta \phi - \frac{3\chi}{4} + L \sqrt{\rho} \sin \left( \frac{\chi}{2}\right)\right] \right\}.
\end{align}

From these equations, we obtain the time evolution of $\Delta \phi = \phi_2 - \phi_1$ as
\begin{align}
\frac{d \Delta\phi}{dt} =& \frac{\Gamma \rho^{1/4} e^{-\sqrt{\rho} L\cos(\chi/2) }}{2\eta \sqrt{2\pi L}} \sin \Delta \phi \nonumber \\
& \times \left\{ -\frac{1}{2L} \cos\left[ -\frac{\chi}{4} + L \sqrt{\rho} \sin \left( \frac{\chi}{2} \right)\right] \right. \nonumber \\
& \left. \quad + \sqrt{\rho} \cos \left[ -\frac{3\chi}{4}  + L \sqrt{\rho} \sin \left(\frac{\chi}{2} \right) \right] \right\},
\end{align}
which corresponds to Eq.~\eqref{eq_pr_s}. In the calculation, we use
\begin{align}
\cos( \Delta \phi + \Xi) - \cos(- \Delta  \phi + \Xi ) = 2 \cos \Xi \sin \Delta \phi 
\end{align}

Next, we consider the case with the opposite rotation direction. In this case, we have to obtain 
$\left. - \nabla u \cdot \bm{e}(-\phi_2 + \pi/2) \right|_{\bm{r} = \bm{r}_2}$.
The polar coordinates corresponding to $\bm{r}_2$ change as $r = \sqrt{L^2 + a^2 - 2 L a \cos \phi_2}$ and $\tan \theta = a \sin \phi_2 / (L - a \cos \phi_2)$. Then, we obtain
\begin{align}
    &\left. -\nabla u^{(1)} \cdot \bm{e}\left(-\phi_2 + \frac{\pi}{2}\right) \right|_{\bm{r} = \bm{r}_2} \nonumber \\
    & \quad= \left. -\frac{\partial u^{(1)}}{\partial r} \bm{e}(\theta) \cdot \bm{e}\left(-\phi_2 + \frac{\pi}{2} \right) \right|_{\bm{r} = \bm{r}_2} \nonumber \\
    & \qquad \left. - \frac{1}{r} \frac{\partial u^{(1)}}{\partial \theta} \bm{e}\left(\theta + \frac{\pi}{2}\right) \cdot \bm{e}\left(-\phi_2 + \frac{\pi}{2} \right)\right|_{\bm{r} = \bm{r}_2} \nonumber \\
    &\quad = \left. \left( \frac{1}{2r} + \sqrt{\rho} \cos \frac{\chi}{2} \right) u^{(1)} \sin(\phi_2+\theta)\right|_{\bm{r} = \bm{r}_2} \nonumber \\
    &\qquad \left. + \left(\frac{1}{r} \cos(\phi_2+\theta)+ \sqrt{\rho} \sin \frac{\chi}{2} \sin(\phi_2+\theta) \right) \hat{u}^{(1)} \right|_{\bm{r} = \bm{r}_2}. \label{force_order1_o}
\end{align}
Equations~\eqref{expand_r}--\eqref{eq_cos} do not change irrespective of the rotation direction within the order of $\mathcal{O}(1)$, and thus we can adopt the expression in Eqs.~\eqref{u_s} and \eqref{utilde_s}. Therefore, we obtain
\begin{align}
&\left. - \nabla u^{(1)} \cdot \bm{e}\left(-\phi_2 + \frac{\pi}{2} \right)\right|_{\bm{r} = \bm{r}_2} \nonumber \\
&\quad = \frac{\rho^{1/4} e^{-\sqrt{\rho} L\cos(\chi/2) }}{2\sqrt{2\pi L}} \nonumber \\
& \qquad \times \left\{
-\frac{1}{2L} \sin \left[-\phi_2 -\phi_1- \frac{\chi}{4} + L \sqrt{\rho} \sin \left(\frac{\chi}{2}\right)  \right] \right. \nonumber \\
& \qquad \left. + \frac{3}{2L} \cos\phi_2 \sin \left[ -\phi_1- \frac{\chi}{4} + L\sqrt{\rho} \sin \left(\frac{\chi}{2}\right)  \right] \right. \nonumber \\
& \qquad \left. + \sqrt{\rho} \sin \phi_2 \cos \left[ -\phi_1- \frac{3\chi}{4} + L \sqrt{\rho} \sin \left(\frac{\chi}{2}\right) \right] \right\} \nonumber \\
& \qquad + \mathcal{O}(a),
\end{align}
Equations~\eqref{aveq_rot1} and \eqref{expansion} give
\begin{align}
& \frac{d\phi_2}{dt} \simeq \omega + \frac{\Gamma \rho^{1/4} e^{-\sqrt{\rho} L\cos(\chi/2) }}{4 \eta \sqrt{2\pi L}} \nonumber \\
& \times \left\{ \frac{3}{2L} \sin \left[ \Delta \phi - \frac{\chi}{4} + L \sqrt{\rho} \sin \left( \frac{\chi}{2}\right) \right] \right. \nonumber \\
& \left. \quad + \sqrt{\rho} \sin \left[\Delta \phi - \frac{3\chi}{4} + L \sqrt{\rho} \sin \left( \frac{\chi}{2}\right)\right] \right\}.
\end{align}
In the same manner as that with the same rotation direction, the time-evolution equation is obtained by considering the geometric symmetry as
\begin{align}
    & \frac{d\phi_1}{dt} \simeq \omega + \frac{\Gamma \rho^{1/4} e^{-\sqrt{\rho} L\cos(\chi/2) }}{4 \eta \sqrt{2\pi L}} \nonumber \\
& \times \left\{\frac{3}{2L} \sin \left[ -\Delta \phi - \frac{\chi}{4} + L \sqrt{\rho} \sin \left( \frac{\chi}{2}\right) \right] \right. \nonumber \\
& \left. \quad + \sqrt{\rho} \sin \left[ - \Delta \phi - \frac{3\chi}{4} + L \sqrt{\rho} \sin \left( \frac{\chi}{2}\right)\right] \right\}.
\end{align}
From these equations, the time-evolution equation for $\Delta \phi$ is obtained as in Eq.~\eqref{eq_pr_o}.

\section{Phase coupling function including the time change in angular velocity\label{app:phase_response_function}}

\begin{figure}
\includegraphics{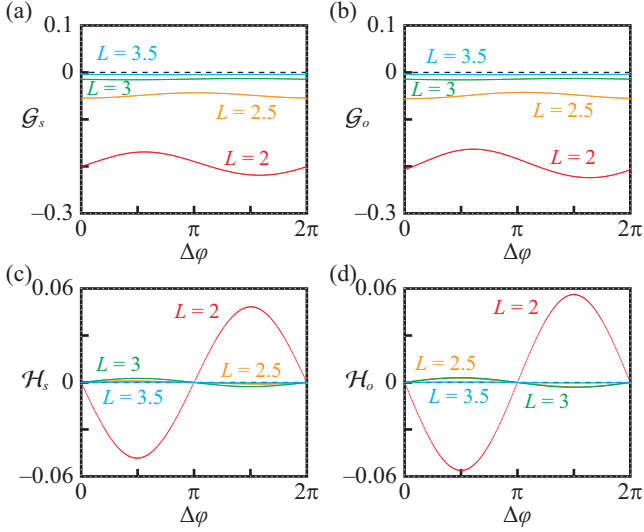}
\caption{\label{fig11} Plots of (a) $\mathcal{G}_s$, (b) $\mathcal{G}_o$, (c) $\mathcal{H}_s$, and (d) $\mathcal{H}_o$ against $\Delta \varphi$. The results with $L = 2$ (red), $L = 2.5$ (orange), $L = 3$ (green), and $L = 3.5$ (cyan) are shown in each panel.}
\end{figure}

\begin{figure}
\includegraphics{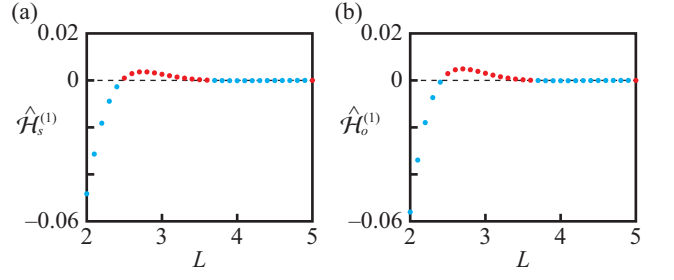}
\caption{\label{fig12} Plots of $\hat{\mathcal{H}}_s^{(1)}$ and $\hat{\mathcal{H}}_o^{(1)}$ against $L$. The positive and negative values are indicated with red and cyan points, respectively. The positive $\hat{\mathcal{H}}_s^{(1)}$ or $\hat{\mathcal{H}}_o^{(1)}$ means that the anti-phase synchronization state is stable, while the negative $\hat{\mathcal{H}}_s^{(1)}$ or $\hat{\mathcal{H}}_o^{(1)}$ means that the in-phase synchronization state is stable.}
\end{figure}

In this section, we discuss the phase coupling function considering the time change in the angular velocity. We calculated the dynamics of one rotor (rotor 1) taking into consideration the concentration field of chemicals generated by the other rotor (rotor 2), though the position of the particle composed of rotor 2 is approximated to be located at the center of it, i.e., $\bm{r}_2(t) = \bm{\ell}_2$, in order to neglect the dependence of $\phi_2$. In this case, the angular velocity of the first rotor $d\phi_1/dt$ is no longer constant but depends on the phase of the rotor 1, $\phi_1$. Thus we have to define a new phase $\varphi_i$, which is defined by $d\varphi_i/dt = 2 \pi / T$, where $T$ is a period. $\varphi_i$ is described as a monotonous increasing function of $\phi_i$. The functions $\mathcal{G}_s(\Delta \varphi)$ and $\mathcal{G}_o(\Delta \varphi)$ are defined as
\begin{align}
  & \mathcal{G}_s(\Delta \varphi) \nonumber \\ &= \frac{1}{2\pi \eta a} \int_0^{2\pi} \bm{e}\left(\varphi_2 - \frac{\pi}{2} \right) \cdot \bm{F}_{u,1,2} (\phi_2 + \Delta \varphi, \varphi_2) d\varphi_2,
\end{align}
\begin{align}
  & \mathcal{G}_o(\Delta \varphi) \nonumber \\ &= \frac{1}{2\pi \eta a} \int_0^{2\pi} \bm{e}\left(-\left(\varphi_2 - \frac{\pi}{2}\right) \right) \cdot \bm{F}_{u,1,2} (\phi_2 + \Delta \varphi, \varphi_2) d\varphi_2 .
\end{align}
Here, we calculated the force $\bm{F}_{u,1,2}$ by taking the rotation of rotor 2 into consideration.
Then, the time-evolution equation of $\Delta \varphi$ is expressed as 
\begin{align}
\frac{d\Delta \varphi}{dt} = \mathcal{G}_i(-\Delta \varphi) - \mathcal{G}_i(\Delta \varphi) \equiv \mathcal{H}_i(\Delta \varphi) 
\end{align}
where $i$ denotes $s$ or $o$.
 The parameters for the simulation were the same as in Sec.~\ref{sec:discussion}. The plots of $\mathcal{G}_s(\Delta \varphi)$, $\mathcal{G}_o(\Delta \varphi)$, $\mathcal{H}_s(\Delta \varphi)$, and $\mathcal{H}_o(\Delta\varphi)$ obtained by the numerical simulation are shown in Fig.~\ref{fig11}. From Figs.~\ref{fig11}(a) and (b), $\mathcal{G}_s(\Delta \varphi)$ and $\mathcal{G}_o(\Delta \varphi)$ are less than those in Figs.~\ref{fig9}(a) and (b). This is due to the interaction through the concentration field, and shows that the effect by the concentration field from the other rotor reduces the averaged angular velocity. Despite the difference in $\mathcal{G}_s(\Delta \varphi)$ and $\mathcal{G}_o(\Delta \varphi)$, $\mathcal{H}_s(\Delta \varphi)$ and $\mathcal{H}_o(\Delta \varphi)$ in Figs.~\ref{fig11}(c) and (d) are almost the same as those in Figs.~\ref{fig9}(c) and (d). 
In the same manner as in the previous paragraph, we consider the Fourier expansion of $\mathcal{H}_s(\Delta \varphi)$ and $\mathcal{H}_o(\Delta \varphi)$ as
\begin{align}
\mathcal{H}_i(\Delta \varphi) = \sum_{k=1}^\infty \hat{\mathcal{H}}_i^{(k)} \sin k \Delta \varphi,
\end{align}
where $i$ denotes $s$ or $o$.
In Fig.~\ref{fig12}, $\hat{\mathcal{H}}^{(1)}_s(\Delta \varphi)$ and $\hat{\mathcal{H}}^{(1)}_o(\Delta \varphi)$ are plotted against $L$, which are almost the same as those in Figs.~\ref{fig7} and \ref{fig10}. This indicates that the decrease in the averaged angular velocity does not matter for the stable synchronization mode, but the interaction between the rotor position and time-dependent concentration field, which is shown in Figs.~\ref{fig8} (c) and (d), plays an important role.

\providecommand{\noopsort}[1]{}\providecommand{\singleletter}[1]{#1}%
%

%\bibliography{paper}

\end{document}